\documentclass[smallextended]{svjour3} 
\smartqed 
\usepackage{graphicx}

\usepackage{amsmath,amssymb}

\numberwithin{equation}{section}

\newcommand{\beq}{\begin{equation}}
\newcommand{\eeq}{\end{equation}}
\newcommand{\be}{\begin{equation}}
\newcommand{\ee}{\end{equation}}
\newcommand{\beqa}{\begin{eqnarray}}
\newcommand{\eeqa}{\end{eqnarray}}
\newcommand{\bea}{\begin{eqnarray}}
\newcommand{\eea}{\end{eqnarray}}

\def\stackunder#1#2{\mathrel{\mathop{#2}\limits_{#1}}}
\newcommand{\comport}[3]{\mathrel{\mathop{#1}\limits_{#2}^{#3}}}

\newcommand{\bino}[2]{\left(#1\atop#2\right)}

\newcommand{\bigmean}[1]{\left\langle#1\right\rangle}
\newcommand{\mean}[1]{\langle#1\rangle}
\newcommand{\bigac}[1]{\left(#1\right)}

\newcommand{\prob}{\mathop{\rm Prob}\nolimits}
\newcommand{\var}{\mathop{\rm Var}\nolimits}

\newcommand{\dd}{{\rm d }}
\newcommand{\ii}{{\rm i}}
\newcommand{\e}{{\rm e}}

\newcommand{\la}{{\lambda}}
\newcommand{\eps}{{\epsilon}}
\renewcommand{\th}{{\theta}}

\renewcommand{\l}{\ell}

\newcommand{\erfc}{\mathop{\rm erfc}}

\renewcommand{\max}{{\rm max}}

\newcommand{\xm}{\langle X\rangle}
\newcommand{\xmtd}{\langle X|L\rangle}
\newcommand{\num}{_{|{\rm num}}}
\newcommand{\lla}{\Lambda}
\newcommand{\td}{\mathop{\scriptstyle \mathrm{td}}}
\newcommand{\free}{\mathop{\scriptstyle \mathrm{f}}}
\newcommand{\pp}{\mathrm{p}}

\newcommand{\cond}{_{|{\rm cond}}}
\newcommand{\dip}{_{|{\rm dip}}}

\journalname{Journal of Statistical Physics}

\begin{document}

\title{Condensation and extremes for a fluctuating number of independent random variables}

\titlerunning{Condensation and extremes} 

\author{Claude Godr\`eche
}

\institute{Claude Godr\`eche \at
Universit\'e Paris-Saclay, CNRS, CEA,
Institut de Physique Th\'eorique,
91191 Gif-sur-Yvette, France\\
\email{claude.godreche@ipht.fr} 
 }

\date{Received: date / Accepted: date}

\maketitle

\begin{abstract}
We address the question of condensation and extremes for three classes of intimately related stochastic processes: (a)
random allocation models and zero-range processes, (b) tied-down renewal processes, (c) free renewal processes.
While for the former class the number of components of the system is fixed, for the two other classes it is a fluctuating quantity.
Studies of these topics are scattered in the literature and usually dressed up in other clothing.
We give a stripped-down account of the subject in the language of sums of independent random variables in order to free ourselves of the consideration of particular models and highlight the essentials.
Besides giving a unified presentation of the theory, this work investigates facets so far unexplored in previous studies.
Specifically, we show how the study of the class of random allocation models and zero-range processes can serve as a backdrop for the study of the two other classes of processes central to the present work---tied-down and free renewal processes. 
We then present new insights on the extreme value statistics of these three classes of processes which allow a deeper understanding of the mechanism of condensation and the quantitative analysis of the fluctuations of the condensate.

\end{abstract}

\section{Introduction}
\label{sec:intro}

It is well known that $n$ independent and identically distributed (iid) positive random variables conditioned by an atypical value of their sum exhibit the phenomenon of condensation, whereby one of the summands dominates upon the others, when their common distribution is subexponential (decaying more slowly than an exponential at large values of its argument).

\begin{figure}[!ht]
\begin{center}
\includegraphics[angle=0,width=.8\linewidth]{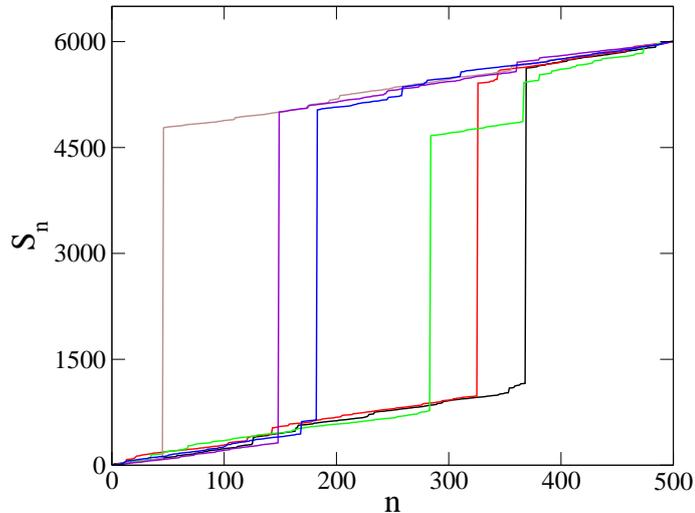}
\caption{Pictorial illustration of the phenomenon of condensation for the random allocation models and \textsc{zrp} class.
This figure depicts six paths of a random walk whose positions are given by the partial sums $S_1,S_2,\dots$.
The distribution of the steps $X_i$ is given by $f(k)=3k^{-5/2}/2$ ($k>1$ is taken continuous), for which 
$\mean{X}=3$.
The random walk is conditioned to end at position $L=6000$ at time $n=500$.
For most of the paths one can observe the occurrence of a big step whose magnitude fluctuates around 
$\Delta=L-\mean{S_n}=4500$.
More rarely, the excess $\Delta$ is shared by two big steps, as for the green path.
}
\label{fig:RW32}
\end{center}
\end{figure}

Let $X_1,X_2,\dots,X_n$ be these $n$ random variables, henceforth taken discrete with positive integer values,
whose common distribution is denoted by $f(k)=\prob(X=k)$.
Hereafter we consider the particular case of a subexponential distribution with asymptotic power-law decay%
\footnote{ In the further course of this work, the symbol $\approx$ stands for asymptotic equivalence;
the symbol $\sim$ means either `of the order of', or
`with exponential accuracy', depending on the context.}
\beq\label{eq:powerlaw}
f(k)\approx \frac{c}{k^{1+\th}},
\eeq
where the index $\th$ and the tail parameter $c$ are both positive.
Assume---for the time being---that the first moment $\xm$ is finite (hence $\th>1$) and that the sum of these random variables,
\be
S_n=\sum_{i=1}^n X_i,
\ee
is conditioned to take the atypically large value $L>\mean{S_n}=n\xm$.
In the present context the phenomenon of condensation has a simple pictorial representation. 
Let us consider the partial sums $S_1,S_2,\dots$ as the successive positions of a random walk whose steps are the summands $X_i$.
Such a representation is used in figure \ref{fig:RW32}, which depicts six different paths of this walk, conditioned by a large, atypical value of its final position $S_n$ after $n$ steps.
The steps have distribution (\ref{eq:powerlaw}) with $\th=3/2$ (see the caption for details).
As can be seen on this figure, for most of the paths there is a single big step bearing the excess difference $\Delta=L-n\xm$.
In the thermodynamic limit, $L\to\infty$, $n\to\infty$ with $\rho=L/n$ fixed, the distribution of the size of this big step becomes narrow around $\Delta$.
This picture gives the gist of the phenomenon of condensation.

The two ingredients responsible for such a phenomenon are: (i) subexponentiality of the distribution $f(k)$, and: (ii) conditioning by an atypically large value of the sum $S_n$.
In contrast, keeping the same distribution (\ref{eq:powerlaw}), but conditioning the sum $S_n$ to be less than or equal to $n\xm$, yields a `democratic' situation where all the steps $X_i$ are on the same footing, sharing the---now negative---excess difference $\Delta$.
Otherwise stated, in this situation, the path responds `elastically' to the conditioning.

When the distribution $f(k)$ is decaying exponentially (e.g., a geometric distribution) the paths responds `elastically' in all cases, i.e., regardless of whether the sum $S_n$ is conditioned to take an atypical value larger or smaller than $n\xm$.
In both cases all the steps $X_i$ are on the same footing, sharing the excess difference $\Delta$ (which is now either positive or negative).

This scenario of condensation has been investigated in great detail and is basically understood.
It is for example encountered in random allocation models where $n$ sites (or boxes) contain altogether $L$ particles, the $X_i$ representing the occupations of these sites \cite{burda,burda2,janson}.
This situation in turn accounts for the stationary state of dynamical urn models such as zero range processes (\textsc{zrp}) or variants \cite{spitzer,andjel,camia,evans2000,jeon,gl2002,cg2003,gross,hanney,gl2005,lux,maj2,ferrari,maj3,armendariz2009,armendariz2011,armendariz2013,armendariz2017,cg2019}.
For short we shall refer to this class of models as the class of random allocation models and \textsc{zrp}.
Condensation means that one of the sites contains a macroscopic fraction of all the particles.

\begin{figure}
\begin{center}
\includegraphics[angle=0,width=.8\linewidth]{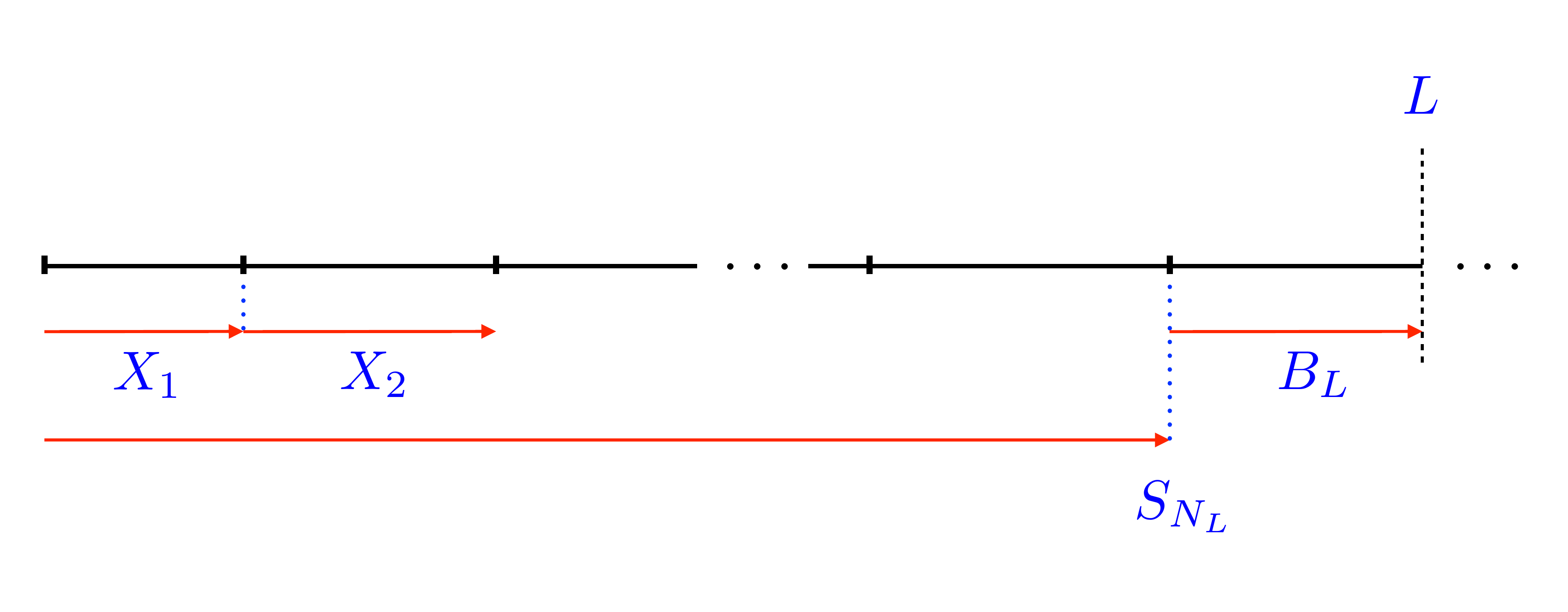}
\caption{In a free renewal process, the number $N_L$ of intervals before $L$ is a random variable such that $S_{N_L}<L<S_{N_{L+1}}$, where $S_{N_L}$ is the sum of the $N_L$ intervals $X_1,X_2,\dots,X_{N_L}$.
The last interval $B_L=L-S_{N_L}$ is unfinished.
For a tied-down renewal process $B_L=0$, $S_{N_L}=L$.
\label{fig:figrenew}}
\end{center}
\end{figure}

Here we shall be concerned by a different situation where the number of random variables $X_1,X_2,\dots,$ is itself a random variable, henceforth denoted by $N_L$ and
defined by conditioning the sum,
\beq\label{eq:SNL}
S_{N_L}=\sum_{i=1}^{N_L} X_i,
\eeq
to satisfy either the inequality
\beq\label{eq:cond1}
S_{N_L}<L<S_{N_L+1},
\eeq
or the equality
\beq\label{eq:cond2}
S_{N_L}=L,
\eeq
where $L$ is a given positive integer number, see figure \ref{fig:figrenew}.
These conditions are imposed irrespectively of whether the mean $\mean{X}$ is finite or not.

The process conditioned by the inequality (\ref{eq:cond1}) defines a \textit{free renewal process} \cite{feller,doob,smith,cox}, the process conditioned by the equality (\ref{eq:cond2}) defines a \textit{tied-down renewal process} \cite{wendel,wendel1,wendel2}.
In the former case the process is pinned at the origin, in the latter case it is also pinned at the end point.
For both, the random variables $X_i$ are the sizes of the iid (spatial or temporal) intervals between two renewals.
 Using the temporal language, the sum
$S_{N_L}$ is the time of occurrence of the last renewal before or at time $L$.
The last, unfinished, interval $B_L=L-S_{N_L}$ is known as the backward recurrence time in renewal theory.
Tied-down renewal processes (\textsc{tdrp}) are special because the pinning condition
(\ref{eq:cond2}) imposes $B_L=0$.

\begin{figure}
\begin{center}
\includegraphics[angle=0,width=0.9\linewidth]{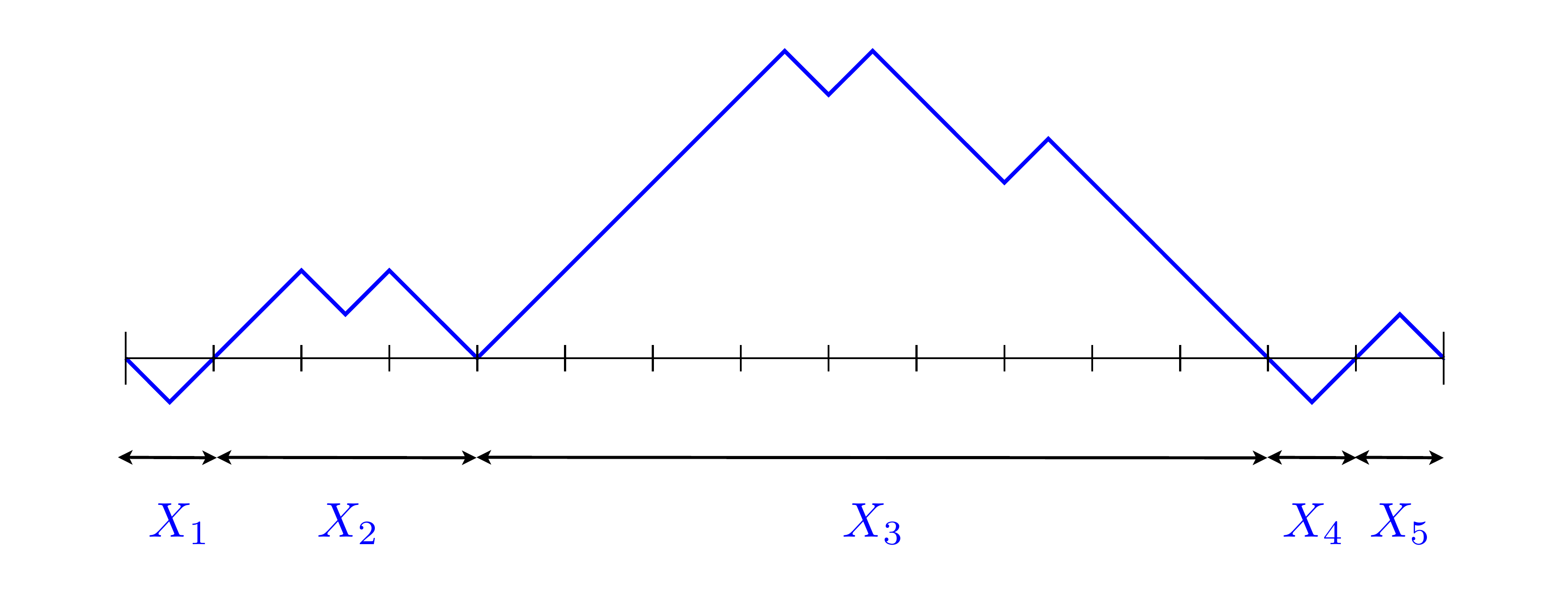}
\caption{A tied-down random walk, or Bernoulli bridge, is a simple random walk, with steps $\pm1$, starting and ending at the origin.
Time is along the $x-$axis, space along the $y-$axis.
The tick marks on the $x-$axis correspond to two time-steps.
In this example the walk is made of $L=15$ tick marks, with $N_{15}=5$ intervals between zeros, $X_1,\dots,X_5$, taking the values $1,3,9,1,1$ tick marks, respectively.
The distribution of the sizes of the intervals, $f(k)=\prob(X=k)$, is given by (\ref{eq:ex1td}).
\label{fig:tdrw}}
\end{center}
\end{figure}

A simple implementation of a \textsc{tdrp} is provided by the Bernoulli bridge, or tied-down random walk,
made of $\pm1$ steps, starting from the origin and ending at the origin at time $2L$ \cite{wendel,wendel1,labarbe}.
The sizes of the intervals between the successive passages by the origin of the walk (where each tick mark on the $x-$axis represents two units of time) represent the random variables $X_i$, as depicted in figure \ref{fig:tdrw}.
The continuum limit of the tied-down random walk is the \textit{Brownian bridge}, also known as \textit{tied-down Brownian motion} or else \textit{pinned Brownian motion} \cite{verv}.

For renewal processes (both free or tied-down) with a subexponential distribution $f(k)$, we shall show that, by weighting the configurations according to the number $N_L$ of summands, a phase transition occurs, as $L\to\infty$, when the positive weight parameter $w$ conjugate to $N_L$, varies from larger values, favouring configurations with a large number of summands, to smaller ones, favouring atypical configurations with a smaller number of summands.
In the former case the weight parameter $w$ is to be interpreted as a reward, in the latter case as a penalty.

Characterising this transition is the aim of the present work,
with main focus on the quantitative analysis of the fluctuations of the condensate.
Again, the occurrence of the phenomenon of condensation is due to: (i) subexponentiality of the distribution $f(k)$, and: (ii) atypicality of the configurations.

Tied-down renewal processes fall into the class of \textit{linear systems} considered by Fisher \cite{fisher}.
The latter are defined as one-dimensional chains of total length $L$, made up, e.g., of alternating intervals of two kinds, $A$ and $B$.
This class encompasses the Poland-Scheraga model \cite{poland,poland2}, consisting of an alternating sequence of straight paths $A$ and loops $B$, wetting models, where $A$ and $B$ represent two phases, etc.
If the direction of the chain is taken as a time axis, the loops in the perpendicular direction can be seen as random walks.
The Bernoulli bridge or tied-down random walk of figure \ref {fig:tdrw} is a natural implementation of this situation, where there is only one kind of intervals, say the loops $B$, representing the intervals between two passages at the origin of the walk.
In the same vein, a variant of the random allocation model defined in \cite{burda} considers the case where the number of sites is varying \cite{burda3}, with an occupation variable $X$ starting at $k=1$.
This model, as well as the spin domain model considered in \cite{bar2,bar3,barma} are examples of linear systems with only one kind of intervals.
Both models are actually equivalent and are just particular instances of the \textsc{tdrp} considered in the present work.
Let us finally mention the random walk (or polymer) models considered in \cite{gia1,gia2} which are free or tied-down renewal processes with a penalty (or reward) at each renewal events, as in the present work.
In these models the condensation transition is interpreted as the transition between a localised phase and a delocalised one \cite{gia1,gia2}.
For the example of figure \ref{fig:tdrw}, a localised configuration corresponds to many contacts of the walk with the origin, while a delocalised one corresponds to the presence of a macroscopic excursion.

Renewal theory is a classic in probability studies.
It is also ubiquitous in statistical physics and has a wide range of applicability (see examples in \cite{gl2001,gms2015,cohen,barkai2003}).
Yet, besides \cite{gia1,gia2}, which give rigorous mathematical results, studies of weighted free renewal processes are scarce.
In particular, there are no available detailed characterisation of the condensation phenomenon for these processes in the existing literature, nor considerations on the statistics of extremes in the condensed phase.

We now describe the organisation of the paper, with further details on the literature on the subject.

The text is composed of three parts, which are respectively sections \ref{sec:zrp} and \ref{sec:zrpLinfini}, dealing with the class of random allocation models and \textsc{zrp}, sections \ref{sec:tdrpGeneral} to \ref{sec:tdrpcondensed}, dealing with \textsc{tdrp}, and finally, sections \ref{sec:rpGeneral} to \ref{sec:rpcondensed}, dealing with free renewal processes.
These three parts are both conceptually and analytically related.

Section \ref{sec:zrp} is a short presentation of how condensation arises for random allocation models and \textsc{zrp}.
Subsection \ref{sec:generalF} gives the basic formalism.
Equation (\ref{eq:qnkL}) introduces a simple remark on the distribution of the maximum in the second half ($k>L/2$), which will turn out to be instrumental in section \ref{sec:zrpLinfini}, and which will be generalised in the later sections on renewal processes.
Subsection \ref{sec:zrpPhenom} 
summarises the main features of the phenomenon of condensation in the thermodynamic limit.
This classical topic has been much investigated in the past, both in statistical physics \cite{burda,burda2,camia,evans2000,gl2002,cg2003,gross,hanney,gl2005,lux,maj2,maj3} and in mathematics \cite{jeon,ferrari,armendariz2009,armendariz2011,armendariz2013,armendariz2017,landim}.
The summary given in this section relies on the short review \cite{cg2019} to which we refer the reader for further bibliographical references.
A mathematical review of a number of aspects of the subject and related matters can be found in \cite{janson}.

Section \ref{sec:zrpLinfini}, which will serve as a backdrop for the study contained in the two other parts, contains novel aspects of the phenomenon of condensation for the class of models at hand.
Namely, instead of considering the thermodynamic limit $L\to\infty$, $n\to\infty$ with fixed ratio $\rho=L/n$, we investigate the situation where the number of summands $n$ is kept fixed and the value of the sum $S_n=L$ increases to infinity.
Such a framework is precisely that considered in \cite{ferrari}. 
In this reference it is shown that for $n$ fixed, $L\to\infty$, assuming that $\rho_c=\xm$ exists, i.e., $\th>1$, if the largest summand is removed, the measure on the remaining summands converges to the product measure with density $\rho_c$, 
a feature which is apparent on figure \ref{fig:RW32}.
Therefore the largest summand is the unique condensate, with size $L-(n-1)\xm$.
As emphasised in \cite{ferrari} `\textit{this phenomenon is a combinatorial fact that can be observed without making the number of sites grow to infinity}'.
Simply stated, there is total condensation in this case, in the sense that the condensate essentially bears the totality of the $L$ particles.
We shall prove this result by elementary means (see (\ref{eq:correctionzrpgt1})) and extend it to the case where $\th<1$, thus proving that even though the first moment $\rho_c=\xm$ does not exist, there is however (total) condensation\footnote{The fact that condensation also occurs when $\th<1$ in the present context has been previously mentioned in \cite{landim}.
I am indebted to S Grosskinsky for pointing this reference to me.}.
If $\th<1$, the correction of the mean largest summand to $L$---in other words the fluctuations of the condensate---scales as $L^{1-\th}$, with a known amplitude, given in (\ref{eq:correctionzrplt1}).

It turns out that this scenario of (total) condensation is precisely that prevailing for the two other processes studied in the present work, namely, tied-down and free renewal processes.
This scenario will be the red thread for the rest of the paper, with all the complications introduced by a now fluctuating number of summands.
This red thread in particular links the three figures \ref{fig:zrp12}, \ref{fig:pp1} and \ref{fig:pp1Free}, and equations (\ref{eq:correctionzrplt1}), (\ref{eq:correctionzrpgt1}),
 (\ref{eq:correctionTdrpLt1}), (\ref{eq:correctionTdrpGt1}), (\ref{eq:correctionRenew}) and (\ref{eq:correctionRenew+}), which are the central results of the present study.

Sections \ref{sec:tdrpGeneral} to \ref{sec:tdrpcondensed} are devoted to \textsc{tdrp} with a power-law distribution of summands (\ref{eq:powerlaw}).
Section \ref{sec:tdrpGeneral} gives a systematic presentation of the formalism, valid for any arbitrary distribution $f(k)$, followed, in sections 
\ref {sec:tdrp-disordered} to \ref{sec:tdrpcondensed} by the analysis of the behaviour of the system in the different phases, when $f(k)$ is a subexponential distribution of the form (\ref{eq:powerlaw}).
As said in the abstract, these topics are scattered in the literature \cite{burda3,bar2,bar3,barma}. 
The analyses presented in sections 
\ref {sec:tdrp-disordered} to \ref{sec:tdrpcondensed} are comprehensive and
go deeper than previous studies, especially in the description of the condensation phenomenon and in the analysis of the statistics of extremes, as detailed in section \ref{sec:tdrpcondensed}.

Likewise, sections \ref{sec:rpGeneral} to \ref{sec:rpcondensed}, devoted to free renewal processes with power-law distribution of summands (\ref{eq:powerlaw}), give a thorough analysis both of the formalism and of the phenomenon of condensation.
It turns out that this case, which is more generic than that of \textsc{tdrp} since the process is not pinned at the end point $L$, is yet more complicated to analyse, the reason being that besides the intervals $X_i$, the last interval $B_L$, depicted in figure \ref{fig:figrenew}, enters the analysis. 

Section \ref{sec:conclusion}, together with tables \ref{tab:tdrp} and \ref{tab:renewal}, gives a summary of the present study.

\section{Condensation for random allocation models and ZRP}
\label{sec:zrp}

The key quantity for the study of the processes described above (random allocation models and \textsc{zrp}, free and tied-down renewal processes) is the statistical weight of a configuration, or in the language of the random walk of figure \ref{fig:RW32}, the statistical weight of a path.

The choice of conventions on the initial values of $n$, $L$, $k$ is a matter of convenience which depends on the kind of reality that we wish to describe, as will appear shortly.

We start 
by giving, in \S\ref{sec:generalF}, some elements of the formalism for random allocation models and \textsc{zrp},
where $f(k)=\prob(X=k)$ is any arbitrary distribution of the positive random variable $X$.
In the rest of section \ref{sec:zrp}, $f(k)$ has a power-law tail (\ref{eq:powerlaw}).

\subsection{General formalism}
\label{sec:generalF}

\subsubsection{Statistical weight of a configuration}
\label{sub:zrpWeight}

Let $X_1,X_2,\dots,X_n$ be $n$ positive iid integer random variables with sum $S_n$ conditioned to be equal to $L$.
The joint conditional probability associated to a configuration $\{X_1=k_1,\dots,X_n=k_n\}$, with $S_n=L$ given, reads\footnote{For the sake of simplicity we restrict the study to the case where $f(k)$ is a normalisable probability distribution.}
\beqa\label{eq:jointezrp}
p(k_1,\dots,k_n|L)&=&\prob(X_1=k_1,\dots,X_n=k_n|S_n=L)
\nonumber\\
&=&\frac{1}{Z_n(L)} f(k_1)\ldots f(k_n)\,\delta\Big(\sum_{i=1}^n k_i,L\Big),
\eeqa
where $\delta(.,.)$ is the Kronecker delta, and where the denominator, whose presence stems from the constraint $S_n=L$, is the partition function
\beqa\label{eq:ZnL}
Z_n(L)&=&\sum_{\{k_i\}} f(k_1)\dots f(k_n)\delta\Big(\sum_{i=1}^n k_i,L\Big)=(f\star)^n(L)
\nonumber\\
&=&\mean{\delta(S_n,L)}=\prob(S_n=L).
\eeqa
So $Z_n(L)$ is another notation for the distribution of the sum $S_n$.

The physical picture associated to these definitions correspond to a system of 
$n$ sites (or boxes), $L$ particles in total, and where the summands $X_i$ are the occupation numbers of these sites, i.e., the number of particles on each of them.
Since these sites can be empty, the occupation probability $f(0)$ is non zero in general.
It is therefore natural to initialise $L$ to 0.
In particular, the probability that the sum of occupations be zero is that all sites are empty, i.e.,
\be
Z_n(0)=f(0)^n.
\ee
It also turns out to be convenient to start $n$ at 0, and set
\beq\label{eq:Z0L}
Z_0(L)=\delta(L,0),
\eeq
which serves as an initial condition for the recursion
\beq\label{eq:recurzrp}
Z_n(L)=\sum_{k=0}^L f(k)Z_{n-1}(L-k).
\eeq
As can be seen either from (\ref{eq:ZnL}) or (\ref{eq:recurzrp}),
the generating function of $Z_n(L)$ with respect to $L$ yields
\beq\label{eq:fgZnL}
\tilde Z_n(z)=\sum_{L\ge0}z^LZ_n(L)=\tilde f(z)^n,
\eeq
where the generating function of $f(k)$ with respect to $k$ is
\be
\tilde f(z)=\sum_{k\ge0}z^k f(k).
\ee
The marginal distribution of the occupation of a generic site, say site $1$, is
\beq\label{eq:zrppkL}
\pi_n(k|L)=\prob(X_1=k|S_n=L)=\mean{\delta( X _1,k)}=\frac{f(k )Z_{n-1}(L-k )}{Z_n(L)},
\eeq
where $\mean{\cdot}$ is the average with respect to (\ref{eq:jointezrp}).
The mean conditional occupation is
\beq\label{eq:xmzrp}
\xmtd=\sum_{k\ge0}k\pi_n(k|L)=\frac{L}{n}=\rho.
\eeq
\subsubsection{Distribution of the largest occupation}
\label{sec:zrpLargest}

Condensation corresponds to the presence of a site with a macroscopic occupation.
We are therefore led to investigate the statistics of the largest occupation.
This topic has been discussed in \cite{janson,jeon,gross,gl2005,maj3,armendariz2009,armendariz2011,cg2019}.
We briefly revisit this topic and supplement it with 
equation (\ref{eq:qnkL}) at the end of this subsection which sheds some new light on the subject and lay the ground for the parallel study of renewal processes.

Let $X_\max$ be the largest summand (or occupation) under the conditioning $S_n=L$,
\be
X_\max=\max(X_1,\dots,X_n).
\ee
The distribution function of this variable is
\beq\label{eq:def-Fn}
 F_n(k|L)=\prob(X_{\max}\le k|S_n=L)
=\frac{\prob(X_{\max}\le k ,S_n=L)}{Z_n(L)},
\eeq
whose numerator is
\beq\label{eq:numeratorzrp}
F_n(k|L)\num=\prob(X_{\max}\le k ,S_n=L)=\sum_{k_1=0}^k f(k_1)\dots \sum_{k_n=0}^k f(k_n)\delta\Big(\sum_{i=1}^n k_i,L\Big).
\eeq
The generating function of the latter reads
\be
\sum_{L\ge0}z^L F_n(k|L)\num
=\prod_{i=1}^n\Big(\sum_{k_i=0}^k f(k_i)z^{k_i}\Big)
=\tilde f(z,k )^n,
\ee
where
\be
\tilde f(z,k )=\sum_{j=0}^k z^{j}f(j).
\ee
The distribution of the largest occupation is thus given by the difference
\be
 p_n(k|L)=\prob(X_{\max}=k|S_n=L)=F_n(k|L)-F_n(k -1|L),
\ee
where
\be
F_n(0|L)\num=f(0)^n\delta(L,0).
\ee
Its generating function is 
\beq\label{eq:p1gen}
\sum_{L\ge0}z^Lp_n(k|L)\num=\tilde f(z,k )^n-\tilde f(z,k -1)^n.
\eeq
The numerator (\ref{eq:numeratorzrp}) obeys the recursion
\be
F_n(k|L)\num=\sum_{j=0}^{\min(k ,L)}f(j)\,F_{n-1}(k|L-j)\num,
\ee
with initial condition
\be
F_{0}(k|L)\num=\delta(L,0).
\ee

Let us note that if the occupation number $X_1$ is larger than $L/2$, then it is necessarily the largest one, $X_\max$.
If so, the probability distribution of the latter, $p_n(k|L)$, is identical to $n\pi_n(k|L)$, since there are $n$ possible choices of the generic summand $X_1$.
Denoting the restriction of $p_n(k|L)$ to the range $k>L/2$ by $q_n(k|L)$, we thus have
\beq\label{eq:qnkL}
q_n(k|L)=n\pi_n(k|L)=\frac{nf(k)Z_{n-1}(L-k)}{Z_n(L)}.
\eeq
We shall see later that this relation, as simple as it may seem, is instrumental for the analysis of the fluctuations of the condensate and extends naturally to the case of \textsc{tdrp} or free renewal processes.
When $k>L/2$ we note that
\beq\label{eq:FnZn}
F_n(k|L-k)\num=Z_n(L-k),
\eeq
since $X_{\max}$ is necessarily less than $k$ (with $k>L/2$) when $S_n=L-k$.

\subsection{Phenomenology of condensation in the thermodynamic limit}
\label{sec:zrpPhenom}

This subsection is a reminder of well-known facts on the phenomenon of condensation for a thermodynamic system with a large number $n$ of sites and 
 large total occupation $L$, at fixed density $\rho=L/n$.
More detailed accounts or complements on this topic can be found, e.g., in 
\cite{burda,burda2,janson,gross,hanney,lux,maj2,maj3,armendariz2009,armendariz2011,cg2019}.
 with further bibliographical references contained in the last reference. 

This reminder will help emphasising the differences between the scenario of condensation \textit{in the thermodynamic limit}, described in the present subsection, with another scenario of condensation, to be described in section \ref{sec:zrpLinfini}, where $L$ is still large, but $n$ (the number of summands or sites) is kept fixed.
In the latter regime, condensation will turn out to be \textit{total}, with a condensed fraction $X_\max/L$ asymptotically equal to unity.

For the time being, we consider the situation where $n$ and $L$ are both large, with $\rho=L/n$ kept fixed, assuming that $f(k)$ has a power-law tail (\ref{eq:powerlaw}) and that $\xm=\rho_c$ is finite ($\th>1$).

\subsubsection{Regimes for the single occupation distribution}

Evidence for the existence of a condensate, i.e., a site with a macroscopic occupation, is demonstrated by the behaviour of the single occupation distribution (\ref{eq:zrppkL}).
There are three regimes to consider, according to the respective values of $\rho$ and $\rho_c$.

\medskip\noindent
\textbf{1. Subcritical regime} ($\rho<\rho_c$)
\\
The asymptotic estimate of the partition function $Z_n(L)$ is given by the saddle-point method
\be
Z_n(L)=\oint\frac{{\rm d}z}{2\pi{\rm i} z^{L+1}}\,\tilde f(z)^n
\sim \frac{\tilde f(z_{0})^n}{z_{0}^{L}},
\ee
where $z_{0}$ obeys the saddle-point (sp) equation
\beq\label{eq:sp}
\frac{z_{0}\tilde f'(z_{0})}{\tilde f(z_{0})}=\rho.
\eeq
This equation has a solution $z_{0}(\rho)$ for any $\rho<\rho_c$.
It follows that
\beq\label{eq:pkLcol}
\pi_n(k|L)_{\rm sp}\approx \frac{z_{0}^kf(k)}{\tilde f(z_{0})},
\eeq
which is no longer dependent on $L$ and $n$ separately and only depends on their ratio $\rho$.
Note that (\ref{eq:sp}) and (\ref{eq:pkLcol}) entail that
\be
\xmtd_{\rm sp}=\sum_{k\ge0}k\pi_n(k|L)_{\rm sp}\approx \frac{z_{0}\tilde f'(z_{0})}{\tilde f(z_{0})}=\rho,
\ee
consistently with (\ref{eq:xmzrp}).
In this regime, the system is made of a fluid of independent particles with common distribution (\ref{eq:pkLcol}).

\medskip\noindent\textbf{2. Critical regime} ($\rho=\rho_c$)
\\
A phase transition occurs when the saddle-point value $z_{0}$ reaches the maximum value of $z$, equal to one, where $\tilde f(z)$ is singular, with a branch cut chosen to be on the negative axis.
The bulk of the partition function is given by the generalised central limit theorem and
\beq\label{eq:zrpcrit}
\pi_n(k|L)\approx f(k),
\eeq
up to finite-size corrections.
At criticality the equality $\xmtd=\xm$ holds identically thanks to (\ref{eq:xmzrp}).
In this regime, the system is made of a critical fluid of independent particles with common distribution (\ref{eq:zrpcrit}).

\medskip\noindent \textbf{3. Supercritical regime} ($\rho>\rho_c$)
\\
In this regime the saddle-point equation (\ref{eq:sp}) can no longer be satisfied because $z_{0}$ sticks to the head of the cut of $\tilde f(z)$.
The excess difference,
\be
\Delta=L-n\xm=n(\rho-\rho_c),
\ee
instead of being equally shared by all the sites, is, with high probability, accommodated by a single site, the \textit{condensate}.
The partition function
$Z_n(L)$ is asymptotically given by its right tail (see 
\cite{cg2019} for more details),
\beq\label{eq:deep}
Z_n(L)\comport{\approx}{n\to\infty}{L=n\rho}\frac{nc}{\Delta^{1+\th}}.
\eeq
In the supercritical regime, the marginal distribution $\pi_n(k|L)$ has different behaviours in the three regions of values of the occupation variable.

\smallskip\noindent \textbf{(a)}
The \textit{critical background} corresponds to values of $k$ finite, for which (\ref{eq:zrpcrit}) holds again.
The main contribution to the total weight comes from this region.

\smallskip\noindent \textbf{(b)} 
The \textit{condensate} is located in the region $k\approx\Delta$ (i.e., the difference $\Delta-k$ is subextensive).
The ratio of $f(k)\approx c/\Delta^{1+\theta}$ to $Z_n(L)$, given by (\ref{eq:deep}), is
asymptotically equal to
\be
\frac{f(k)}{Z_n(L)}\approx \frac{c/\Delta^{1/\theta}}{nc/\Delta^{1/\theta}}=\frac{1}{n}.
\ee
On the other hand, $Z_{n-1}(L-k)$, is given by its bulk since $L-k\approx n\rho_c$.
Hence, if $1<\theta<2$,
\beq\label{eq:bulkL}
\pi_n(k|L)\cond\approx\frac{1}{n}Z_{n-1}(L-k)\approx\frac{1}{n}\frac{1}{n^{1/\theta}}\mathcal{L}_{\theta,c}\left(\frac{\Delta-k}{n^{1/\theta}}\right),
\eeq
where $\mathcal{L}_{\theta,c}$ is the stable L\'evy distribution of index $\th$, asymmetry parameter $\beta=1$, and tail parameter $c$ \cite{gnedenkoK}, 
while, if $\theta>2$,
\beq\label{eq:bulkG}
\pi_n(k|L)\cond\approx\frac{1}{n}Z_{n-1}(L-k)\approx\frac{1}{n}\frac{1}{n^{1/2}}\mathcal{G}\left(\frac{\Delta-k}{n^{1/2}}\right),
\eeq
where $\mathcal{G}$ is the Gaussian distribution \cite{gnedenkoK}.
These expressions describe the bulk of the fluctuating condensate which manifests itself by a hump in the marginal distribution $\pi_n(k|L)$, in the neighbourhood of $k\approx\Delta$, visible on figure \ref{fig:zrp-marg-max}.
The weight of this region is obtained from (\ref{eq:bulkL}) or (\ref{eq:bulkG}), according to the value of $\th$, as
\beq\label{eq:weightzrphump}
\prob(X_1\in \mathrm{cond})=\sum_{k\in\mathrm{hump}}\pi_n(k|L)\cond\approx
\frac{1}{n},
\eeq
which demonstrates that the excess difference $\Delta$ is typically borne by only one summand.

This hump becomes peaked in the thermodynamic limit.
For a finite system, most often there is a single condensate, i.e., a site with a macroscopic occupation, while more rarely there are two sites with macroscopic occupations, both of order $L$.
This situation corresponds to the \textit{dip region}, described next.

\smallskip\noindent\textbf{(c)} 
 The range of values of $k $ such that $k$ and $\Delta-k$ are large and comparable, interpolates between the critical part of $ \pi_n(k|L)$, for $k $ or order $1$, and the condensate, for $k $ close to $\Delta$.
It corresponds to the \textit{dip region} on figure \ref{fig:zrp-marg-max}.
In this region, $Z_{n-1}(L-k )$ is given by its right tail (\ref{eq:deep}).
So, for any $\theta>1$,
\beq\label{eq:dipzrp}
\pi_n(k|L)\dip
 \approx 
 c\left[\frac{\Delta}{k (\Delta-k )}\right]^{1+\theta}
 \approx 
 \frac{f(k )f(\Delta-k )}{f(\Delta)}.
\eeq
The interpretation of this result is that in the dip region
typical configurations where one summand takes the value $k $ 
are such that the remaining $\Delta-k $ excess difference
is borne by a single other summand.
The dip region is therefore dominated
by rare configurations where the excess difference is shared by {\it two} summands \cite{gl2005}.
An example of such a configuration is the green path in figure \ref{fig:RW32}.

Setting $k=\la \Delta$ in (\ref{eq:dipzrp}) and introducing a cutoff $\lla=\eps \Delta$, the weight of these configurations can be estimated as 
\beq\label{eq:zrpWeightdip}
\prob(X_1\in \mathrm{dip})=\sum_{k=\eps\Delta}^{(1-\eps)\Delta} \pi_n(k|L)\dip
 \sim \Delta^{-\theta}\sim n^{-\theta}.
\eeq
The relative weights of the dip (\ref{eq:zrpWeightdip}) and condensate (\ref{eq:weightzrphump}) regions is therefore of order $n^{-(\theta-1)}$,
i.e., the weight of events where the condensate is broken into two pieces of order $n$ is subleading with respect to events with a single big summand.

\subsubsection{Statistics of the largest summand in the condensed phase}
\label{sec:zrpMaxim}

In view of (\ref{eq:weightzrphump}) and the following equations (\ref{eq:dipzrp}) and (\ref{eq:zrpWeightdip}), we infer that, for $k$ larger than $\Delta/2$, which is the centre of the dip,
the excess difference $\Delta$ is typically borne by only one summand---namely the condensate $X_\max$---thus
\beq\label{eq:condPi}
 p_n(k|L)\stackunder{k>\Delta/2}{\approx} n\pi_n(k|L).
 \eeq
According to (\ref{eq:qnkL}) we know that the two sides of this equation are actually identical for $k>L/2$, for any finite values of $n$ and $L$.
Note however that, while $L/2$ is always larger than $\Delta/2$, it can be smaller or larger than $\Delta$ depending on whether $\rho$ is larger or smaller than $2\rho_c$.
The significance of (\ref{eq:condPi}) is that the equality (\ref{eq:qnkL}), valid for $k>L/2$, extends asymptotically to the entire region $k>\Delta/2$. 

More precise statements have been given 
on the asymptotic distribution of the largest summand $X_\max$ in
\cite{maj3,armendariz2009,armendariz2011,janson}. 
The result is that, if $L=n\rho$, $\rho>\rho_c$, $n\to\infty$, 
the rescaled variable
$n^{-1/\alpha}(\Delta-X_\max)$
converges to a stable law of index $\alpha$, with $\alpha=\theta$ if $\theta<2$, or $\alpha=2$ if $\theta>2$.
This means that, asymptotically, the probability distribution of $X_\max$ coincides, up to a factor $n$, with the estimates of the marginal density in the condensate region ($\Delta-k\sim n^{1/\alpha}$), that is with (\ref{eq:bulkL}) or (\ref{eq:bulkG}) according to the value of $\theta$, which is precisely the content of (\ref{eq:condPi}).

On the other hand,
denoting the $r-$th largest summand by $X^{(r)}$ ($r=1,\dots,n$), with $X^{(1)}\equiv X_\max$,
the distributions of these ranked summands, denoted by $p_n^{(r)}(k|L)$ with $p_n^{(1)}(k|L)\equiv p_n(k|L)$, sum up exactly to
\beq\label{eq:rank}
p_n(k|L)+\sum_{r=2}^n p_n^{(r)}(k|L)=n\pi_n(k|L).
\eeq
Thus according to (\ref{eq:condPi}) the sum upon $r\ge2$ in the left side of (\ref{eq:rank}) is negligible for $k>\Delta/2$.

As shown in \cite{armendariz2009,armendariz2011,janson},
the distribution of the second largest summand, $X^{(2)}$, is asymptotically Fr\'echet, and the subsequent ones, 
$X^{(r)}$ 
$(r\ge 2)$, are the order statistics of $n-1$ iid random variables $X_i$ with distribution $f(k)$,
which amounts to saying that, in the supercritical regime, the dependency between the summands $X_i$ introduced by the conditioning goes asymptotically in the condensate $X_\max$.

Since $X_\max$ typically scales as $n$, while $X^{(2)}, X^{(3)},\dots$ typically scale as $n^{1/\theta}$,
the condensate is increasingly separated from the background as $n$ increases, leaving space to the dip region ($k$ and $\Delta-k$ large and comparable).
We know from the analysis made above (see discussion following (\ref{eq:dipzrp})) that 
this region is dominated
by configurations where the excess difference is shared by two summands, namely $X_\max$ and $X^{(2)}$, so 
\beq\label{eq:f12dip}
 p_n(k|L)+p_n^{(2)}(k|L) \stackunder{k\in \mathrm{dip}}{\approx} n\pi_n(k|L) |_{\rm dip},
\eeq
and that the contributions of these events to $n\pi_n(k|L)$ are of relative order $n^{-(\theta-1)}$.
To the right of $\Delta/2$ the predominant contribution to the sum on the left side of (\ref{eq:f12dip}) comes from $p_n(k|L)$, to the left of $\Delta/2$ it comes from
$p_n^{(2)}(k|L)$.

\subsubsection*{An illustration}

\begin{figure}[!ht]
\begin{center}
\includegraphics[angle=0,width=.8\linewidth]{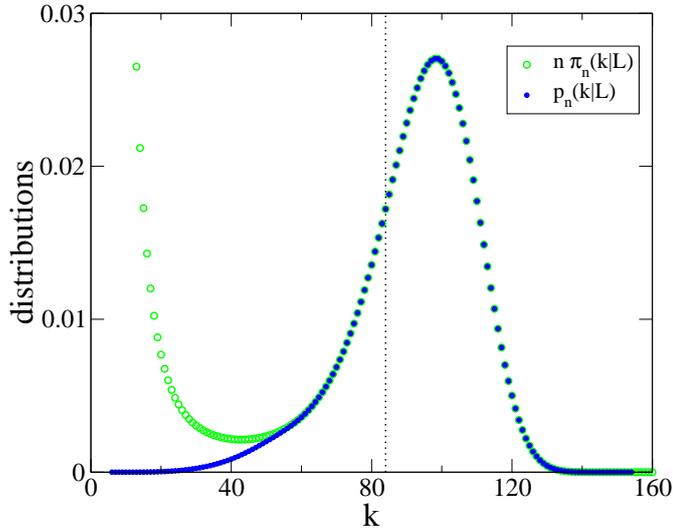}
\caption{Random allocations models and \textsc{zrp}: comparison of $n\pi_n(k|L)$ (where $\pi_n(k|L)$ is the single occupation distribution) with $p_n(k|L)$ (distribution of the maximum) for the example (\ref{eq:zeta}), with $\th=3$,
$\rho_c\approx0.1106$, $n=600$, $L=168$, $\Delta\approx 102$.
The vertical dotted line is at $L/2$.
There is an exact identity between $n\pi_n(k|L)$ and $p_n(k|L)$, for $k>L/2$, as explained in \S \ref {sec:zrpLargest}.
Moreover, there is already excellent numerical coincidence between the two curves as soon as 
$k>58\gtrsim\Delta/2$.
}
\label{fig:zrp-marg-max}
\end{center}
\end{figure}

Figure \ref{fig:zrp-marg-max} depicts a comparison between $n\pi_n(k|L)$, obtained from (\ref{eq:zrppkL}), and $p_n(k|L)$ obtained from (\ref{eq:p1gen}),
on the following example, defined by the normalised distribution
\beq\label{eq:zeta}
f(k)=\frac{1}{\zeta(1+\th)}\frac{1}{(k+1)^{1+\th}}, \quad (k\ge0),
\eeq
where $\zeta(s)=\sum_{s\ge1}1/n^s$ is the Riemann zeta function.
This model has been introduced in \cite{burda}, then further investigated in \cite{camia,burda2,glzeta}.

In this figure, $\th=3$, $\rho_c\approx 0.1106$, $n=600$, $L=168$.
This choice of parameters corresponds to a density $\rho=0.28$ slightly larger than $2\rho_c$, where $\Delta$ and $L/2$ coincide.
The top of the hump is approximately located at $\Delta\approx 102$ and the minimum of the dip is a bit less than $\Delta/2$.
These curves are practically indiscernible as soon as $k\approx 58$, which is less than $L/2=84$, indicated by the vertical dotted line on the figure, from which the identity becomes exact.

\begin{figure}[!h]
\begin{center}
\includegraphics[angle=0,width=.8\linewidth]{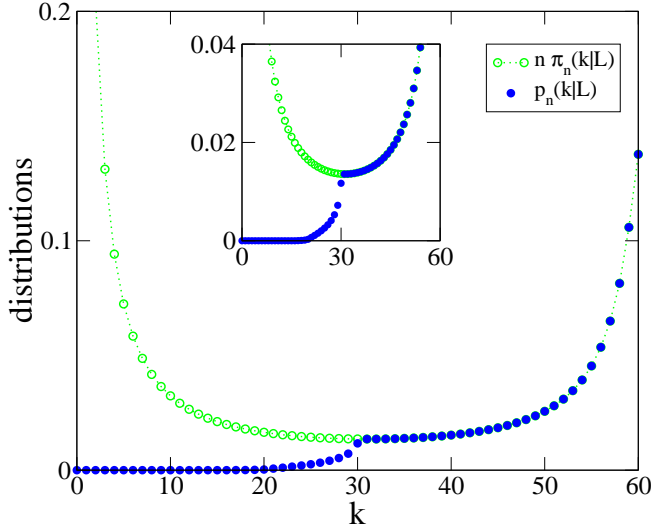}
\caption{Random allocations models and \textsc{zrp}: comparison of $n\pi_n(k|L)$ (where $\pi_n(k|L)$ is the single occupation distribution) with $p_n(k|L)$ (distribution of the maximum) for the example (\ref {eq:zrp12}) corresponding to a tail index $\th=1/2$, with $L=60$, $n=4$.
The inset highlights the cusp at $L/2$.
}
\label{fig:zrp12}
\end{center}
\end{figure}

Finally, let us compute the mean condensed fraction $X_\max/L$.
For $\th>2$, using (\ref{eq:bulkG}), it can be estimated as
\bea 
\frac{\mean{X_\max}}{L}&=&\frac{1}{L}\sum_{k=0}^L k\, p_n(k|L)
\approx \frac{1}{L}\sum_{k=\Delta/2}^L k\, n\,\pi_n(k|L)
\nonumber\\
&\approx&\frac{1}{L}\frac{1}{n^{1/2}\sqrt{2\pi}}\int_{-\infty}^\infty\dd k\, k\,\e^{-\frac{1}{2}
\left(\frac{\Delta-k}{n^{1/2}}\right)^2}=1-\frac{\rho_c}{\rho}=\frac{\Delta}{L}.
\eea
The same result holds if $1<\th<2$, using (\ref{eq:bulkL}).

As $\rho$ increases, the peak of the condensate moves towards the right end $L$, hence if $\rho\gg \rho_c$ the condensed fraction tends to unity, corresponding to total condensation.
As detailed in section \ref{sec:zrpLinfini}, this scenario still holds when $n$ is kept fixed.

\section{Phenomenon of total condensation when $n$ is kept fixed and $L\to\infty$}
\label{sec:zrpLinfini}

As seen above, if $\rho\gg \rho_c$, condensation becomes total, and the peak of the condensate is asymptotically located at $L$.
As we now show, this still holds true if the number of summands $n$ is kept fixed, and $L$ is large, 
irrespective of the existence of a first moment $\rho_c=\xm$, or in other words, irrespective of whether 
$\th$ is smaller or larger than one.
Existence of condensation in such a situation has been pointed out in \cite{ferrari,landim}.
The quantitative characterisation of this phenomenon is the aim of this section.

\subsection{An illustration}
We start by giving an illustration of the phenomenon on the following example, corresponding to a tail index $\th=1/2$ for the decay of $f(k)$, 
\beq\label{eq:zrp12}
\tilde f(z)=\frac{1-\sqrt{1-z}}{z},
\eeq
which entails that
\be
f(k)=\frac{1}{2^{2k+1}}\frac{(2k)!}{k!(k+1)!}
\stackunder{k\to\infty}{\approx} \frac{c}{ k ^{3/2}},
\ee
with $c=1/(2\sqrt{\pi})$,
and
\beq\label{eq:ZnLnfixed}
Z_n(L)=\frac{n}{2^{2L+n}(2L+n)}\bino{2L+n}{L}\comport{\approx}{L\to\infty}{n\ \mathrm{ fixed}} \frac{nc}{L^{3/2}},
\eeq
for $n$ kept fixed, $L$ large, which is a different regime from that leading to (\ref {eq:deep}).
In the general case, whenever $f(k)$ obeys (\ref{eq:powerlaw}), the asymptotic estimate (\ref{eq:ZnLnfixed}) becomes
\beq\label{eq:ZnLnfixed+}
Z_n(L)\comport{\approx}{L\to\infty}{n\ \mathrm{ fixed}} \frac{nc}{L^{1+\th}},
\eeq
as can be deduced from (\ref{eq:fgZnL}). 

Figure \ref{fig:zrp12} depicts a
comparison of $n\pi_n(k|L)$ with $p_n(k|L)$ for this example with $L=60$, $n=4$.
These two quantities are identical for $k>L/2$, 
The inset highlights the existence of a cusp at $L/2$.
The distribution of the maximum $p_n(k|L)$ is significantly depressed for $k<L/2$.
It vanishes identically for $k<L/4$ (more generally $L/n$).
These features are analysed in the two subsections below.

\subsection{Fluctuations of the condensate}
\label{sec:zrpFluctu}
Let us turn to the general case.
A measure of the fluctuations of the condensate is provided by 
the width of the peak of the maximum, i.e., the mass outside the condensate.
It can be estimated by the sum (see \S\ref{sec:details} below for details)
\beq\label{ZRP-width}
 L-\mean{X_\max}=\sum_{\l=0}^{L-1}\l\, p_n(L-\l|L)
\approx \sum_{\l=0}^{L/2-1}\l\, q_n(L-\l|L)
\approx \sum_{\l=0}^{L/2-1}\l\, n\pi_n(L-\l|L),
\eeq
where $\l=L-k$.
The dominant contribution to this sum depends on whether $\th$ is smaller or larger than one.

\vspace{4pt}
\smallskip\noindent$\bullet$
 If $\th<1$, the dominant contribution comes from values of $\l=L-k$ comparable to $L$.
Setting $\l=\lambda L$ in (\ref{ZRP-width}), we have (see \S\ref{sec:details} below for details)
\beq\label{eq:correctionzrplt1}
 L-\mean{X_\max}
\approx (n-1)c L^{1-\th}\int_{0}^{1/2}\dd \la \frac{\la}{[(\la(1-\la)]^{1+\th}}
\approx (n-1)c\,\mathrm{B}_{\frac{1}{2}}\Big(1-\th,-\th\Big)L^{1-\th},
\eeq
where the incomplete beta function is defined as
\be
\mathrm{B}_{x}(a,b)=\int_0^x\dd t\,t^{a-1}(1-t)^{b-1}.
\ee
For example, if $\th=1/2$, we have $\mathrm{B}_{\frac{1}{2}}\Big(1-\th,-\th\Big)=2$.

\vspace{4pt}
\smallskip\noindent $\bullet$
If $\th>1$, the main contribution comes from finite values of $\l$, 
\beqa\label{eq:correctionzrpgt1}
 L-\mean{X_\max}
&\approx& n\sum_{\l=0}^{L/2}\l\, \frac{f(L-\l)Z_{n-1}(\l)}{Z_n(L)}
\approx \sum_{\l=0}^{L/2}\l\, Z_{n-1}(\l)
\nonumber\\
 &\approx& (n-1) \sum_{\l=0}^{L/2}\l\, f(\l)
\to (n-1) \sum_{\l=0}^{\infty}\l f(\l)
= (n-1)\xm.
\eeqa
This last result (\ref{eq:correctionzrpgt1}) has a simple interpretation.
It says that the correction $L-\mean{X_\max}$ comes from the $n-1$ sites of the fluid,
each with mean occupation $\xm=\rho_c$, in accordance with the prediction made in \cite{ferrari} and recalled in the introduction.

\subsection{A finer analysis}
\label{sec:details}
Let us now add some more details on the derivations made above.
The aim of this subsection is to give a detailed analysis of the distributions in the various regimes, in order to eventually compute the corrections to the scaling expressions predicted in (\ref{eq:correctionzrplt1}) and (\ref{eq:correctionzrpgt1}) above.

We start with the discussion of the regimes for the single occupation distribution $\prob(X_1=k|S_n=L)=\pi_n(k|L)$.
There are such three regimes to consider (see figure \ref{fig:zrp12}):
\begin{enumerate}
\item \textit{Downhill region}.
 For $X_1=k$ finite, using (\ref{eq:ZnLnfixed+}), we have
\be
\pi_n(k|L)\stackunder{L\to\infty}{\approx} \frac{n-1}{n}f(k),
\ee
reflecting the fact that, with probability $(n-1)/n$, a randomly chosen site belongs to the fluid.

Introducing a cutoff $\lla$, such that $1\ll\lla\ll L$, the weight of this region can be estimated by the sum
\beq\label{eq:region1ZRP}
\sum_{k=0}^\lla \pi_n(k|L) \approx \sum_{k=0}^\lla \frac{n-1}{n}f(k)\stackunder{L\to\infty}{\to} 
\sum_{k=0}^\infty \frac{n-1}{n}f(k)=
1-\frac{1}{n}.
\eeq

\item
\textit{Dip region}.
In the dip region, where $k$ and $L-k$ are simultaneously large and comparable, setting $k=\la L$ in (\ref{eq:zrppkL}) where $0<\la<1$,
and using (\ref{eq:ZnLnfixed+}), yields the estimate
\beq\label{eq:piDip}
\pi_n(k|L)\stackunder{L\to\infty}{\approx}\frac{n-1}{n} \frac{f(k)f(L-k)}{f(L)}\approx
\frac{n-1}{n}\frac{c}{L^{1+\th}}\frac{1}{[\la(1-\la)]^{1+\th}},
\eeq
In this region the distribution is therefore U-shaped: the most probable configurations are those where almost all the particles are located on one of two sites. 
The dip centred around $k=L/2$ becomes deeper and deeper with $L$.

The weight of this region reads, choosing $\lla=\eps L$,
\beq\label{zrpDip}
\sum_{k=\lla}^{L-\lla}\pi_n(k|L)\approx \frac{n-1}{n}c L^{-\th}
\int_\eps^{1-\eps}\frac{\dd\la}{[\la(1-\la)]^{1+\th}}.
\eeq

\item \textit{Uphill region}.
The condensate region corresponds to $\l=L-k$ finite, where
 (\ref{eq:zrppkL}) simplifies into
\beq\label{eq:piUphill}
\pi_n(L-\l|L)\stackunder{L\to\infty}{\approx} \frac{1}{n}Z_{n-1}(\l),
\eeq
as in (\ref{eq:bulkL}) or (\ref{eq:bulkG}).
The weight of this uphill region can be estimated as
\beq\label{eq:region2ZRP}
\sum_{\l=0}^\lla \pi_n(L-\l|L)\approx\frac{1}{n}\sum_{\l=0}^\lla Z_{n-1}(\l)
\to\frac{1}{n}\sum_{\l=0}^\infty Z_{n-1}(\l)=\frac{1}{n},
\eeq
where $\lla$ is yet another cutoff, and where the last step is obtained by setting $z=1$ in the expression of the generating function (\ref{eq:fgZnL}).
Thus, as seen in (\ref{eq:region1ZRP}) and (\ref{eq:region2ZRP}) 
the weights of the downhill and uphill regions add up to one, in line with the fact that the contribution of the dip region is subdominant, as shown in (\ref{zrpDip}) above.

\end{enumerate}

We now proceed to the discussion of the regimes for the distribution of the maximum, $p_n(k|L)$.
There are again three regimes to consider, that we describe in turn, from right to left in figure \ref{fig:zrp12}.
In the uphill and dip regions, such that $X_\max=k>L/2$, $p_n(k|L)$ is denoted by $q _n(k|L)=n\pi_n(k|L)$ (see (\ref{eq:qnkL})), whose estimates follow from those of $\pi_n(k|L)$ seen above.

\begin{enumerate}
\item \textit{Uphill region.} For $\l$ finite,
using (\ref{eq:piUphill}), we have
\beq\label{eq:uphill-cond}
q_n(L-\l|L)\approx Z_{n-1}(\l),
\eeq
with weight equal to 1 up to the subleading corrections detailed below.
The interpretation of (\ref{eq:uphill-cond}) is that, asymptotically, the difference between $L$ and $X_\max$ has the same distribution as the sum of $n-1$ iid random variables, the latter composing the fluid,
\beq\label{eq:uphill-subexp}
L-X_\max\stackunder{L\to\infty}{\approx} \
\stackunder{\rm fluid}{\underbrace{\sum_{i=1}^{n-1}X_{i}}}.
\eeq
\item 
\textit{Dip region.}
For $L/2<k\sim L-k$,
we have, according to (\ref{eq:piDip}),
\beq\label{eq:zrpMaxDip}
q_n(k|L)\approx \frac{(n-1)f(k)f(L-k)}{f(L)}\approx \frac{(n-1)\,c}{L^{1+\th}}
\frac{1}{[\la(1-\la)]^{1+\th}}.
\eeq
The weight of this region therefore scales as $L^{-\th}$, as seen in (\ref{zrpDip}).

\item \textit{Left region.}
For $k\le L/2$, the weight of this region is subdominant with respect to that of the two previous ones.
A simple argument shows that
\beq\label{eq:FnLsur2}
\prob(X_\max\le L/2|S_n=L)=F_n(L/2|L)\sim L^{-\beta},\qquad 
\beta= \left\{
\begin{array}{ll}
2\th \ & \textrm{if } \th\le1
\vspace{4pt}\\
1+\th\ & \textrm{if }\th>1,
\end{array}
\right.
\eeq 
where $F_n$ is defined in (\ref{eq:def-Fn}). 
\end{enumerate}
The argument leading to (\ref{eq:FnLsur2}) and the prediction of the amplitude 
$\lim_{L\to\infty}F_n(L/2|L)L^{\beta}$ are given in appendix \ref{app:heuristic}.
In appendix \ref{app:levy}, we give an exact calculation of the weight of the left region when $f(k)$ is the continuous L\'evy $\frac{1}{2}$ stable density.

Equation (\ref{eq:FnLsur2}) eventually justifies the approximation made in (\ref{ZRP-width}), where the contribution of the left region to the sum $\sum_{\l=0}^{L-1}\l\, p_n(L-\l|L)$ was neglected.
In view of (\ref{eq:FnLsur2}), this contribution is $O(L^{1-\beta})$, thus subdominant by a factor $L^{-\th}$ with respect to the first correction---respectively $O(L^{1-\th})$ if $\th<1$ (see (\ref{eq:correctionzrplt1})), or $O(1)$ for $\th>1$ (see (\ref{eq:correctionzrpgt1}))---whether $\th$ is smaller of larger than unity.

There is actually a hierarchy of weights for the distribution of the maximum in the successive regions $(L/3,L/2)$, $(L/4,L/3)$ and so on.
This can be intuitively grasped as follows.
\begin{enumerate}
\item If $L/3< X_\max\le L/2$, the total `mass' $L$ is dominantly shared by two summands.
The weight of this rare event scales as in (\ref{eq:FnLsur2}).
\item
Then if $L/4< X_\max\le L/3$, the total `mass' $L$ is dominantly shared by three summands, which is a still rarer event,
and so on.
\item
Finally, if $X_\max<L/n$, the probability $p_n(k|L)$ vanishes since it is no longer possible to divide the `mass' $L$ into $n$ pieces, all less than $L/n$.
\end{enumerate}
This hierarchy is reflected by the presence of cusps in the distribution of the maximum at $L/2, L/3$, $L/4\dots$ (see appendix \ref{app:heuristic}).

All the discussion given in the present section is a preparation for subsections \ref{sub:regimesMax} and \ref{sub:RPlongest}, where figures \ref{fig:pp1} and \ref{fig:pp1Free} are to be compared to figure \ref{fig:zrp12},
and equations 
 (\ref{eq:correctionTdrpLt1}), (\ref{eq:correctionTdrpGt1}), 
 (\ref{eq:correctionRenew}) 
 and (\ref{eq:correctionRenew+}) are to be compared 
to equations (\ref{eq:correctionzrplt1}) and (\ref{eq:correctionzrpgt1}).

\section{General statements on tied-down renewal processes}
\label{sec:tdrpGeneral}
The random variables $X_i$ now represent the sizes of (spatial or temporal) intervals, that we take strictly positive, hence 
\beq\label{eq:f0}
f(0)=0.
\eeq
In the temporal language $L$ is the total duration of the process, in the spatial language it is the length of the system.
To each interval (equivalently, to each renewal event) is associated a positive weight $w$, to be interpreted as a reward if $w>1$ or a penalty if $w<1$.
In the models considered in \cite{burda3,bar2,bar3,barma}, $w$ has the interpretation of the ratio $y/y_c$, where $y$ is a fugacity, and $y_c$ its value at criticality.

In the present section, $f(k)=\prob(X=k)$ is any arbitrary distribution of the positive random variable $X$.
Later on, in sections \ref{sec:regimetdrp}, 
\ref{sec:tdrp-disordered}, \ref{sec:tdrp-critical} and \ref{sec:tdrpcondensed}, $f(k)$ will obey the form (\ref{eq:powerlaw}).

\subsection{Joint distribution}

The probability of the configuration $\{X_1=k_1,\dots,X_{N_L}=k_n,N_L=n\}$, given that $S_{N_L}=L$, reads
\beqa\label{eq:jointe}
 p(k_1,\dots,k_n,n|L)
&=&\prob(X_1=k_1,\dots,X_{N_L}=k_n,N_L=n|S_{N_L}=L)
\nonumber\\
 &=&\frac{1}{Z^{\td}(w,L)}w^nf(k_1)\dots f(k_n)\delta\Big(\sum_{i=1}^n k_i,L\Big),
\eeqa
where the denominator is the tied-down partition function
\beqa\label{eq:ZwL}
Z^{\td}(w,L)
&=&\sum_{n\ge0}w^n\sum_{\{k_i\}} f(k_1)\dots f(k_n)\delta\Big(\sum_{i=1}^n k_i,L\Big)
\nonumber\\
&=&\sum_{n\ge0}w^nZ_n(L)
=\delta(L,0)+\sum_{n\ge1}w^n(f\star)^{n}(L).
\eeqa
The probability $Z_n(L)$ is still defined as in (\ref{eq:ZnL}), except for the change of the initial value (\ref{eq:f0}) of $f(k)$, which entails that $Z_n(L)$ is only defined for $n\le L$.
The first term $\delta(L,0)$ follows from (\ref{eq:Z0L}).
The first values of $Z^{\td}(w,L)$ are
\bea
 Z^{\td}(w,0)=1,\quad Z^{\td}(w,1)=wf(1),\quad Z^{\td}(w,2)=w f(2)+w^2f(1)^2,
\nonumber\\
 Z^{\td}(w,3)=wf(3)+2w^2f(1)f(2)+w^3f(1)^3,
\eea
and so on.
The generating function of $Z^{\td}(w,L)$ with respect to $L$ is
\beq\label{eq:fgZ}
\tilde Z^{\td}(w,z)=\sum_{L\ge0}z^L Z^{\td}(w,L)=\sum_{n\ge0}w^n\tilde f(z)^n=
\frac{1}{1-w\tilde f(z)}.
\eeq
Note that $Z^{\td}(w,L)$ can be seen as the grand canonical partition function of the system with respect to $N_L$.

For $w=1$, the tied-down partition function,
\beq\label{eq:Z1L}
 Z^{\td}(1,L)=\sum_{n\ge0}\prob(S_n=L)=\prob(S_{N_L}=L)=\mean{\delta(S_{N_L},L)},
\eeq
is the probability that a renewal occurs at $L$.

Finally, we note that
\be
p(k_1,\dots,k_n,n|L)\num=w^n p_n(k_1,\dots,k_n|L)\num,
\ee
hence
\beq\label{eq:Newppnpn}
p(k_1,\dots,k_n,n|L)=p_n(k_1,\dots,k_n|L)\pp_n(L),
\eeq
where $p_n(\{k_i\}|L)$ and $\pp_n(L)$ are respectively defined in (\ref{eq:jointezrp}) and (\ref{eq:pnL}).

\subsection{Distribution of the number of intervals}
The distribution of the number of intervals is obtained by summing the distribution (\ref{eq:jointe}) upon all variables except $n$, 
\beq\label{eq:pnL}
\pp_n(L)=\prob(N_L=n)=\frac{w^nZ_n(L)}{Z^{\td}(w,L)}.
\eeq
For instance, taking the successive terms of (\ref{eq:ZwL}) divided by $Z^{\td}(w,L)$ yields
\be
 \pp_0(L)=\frac{\delta(L,0)}{Z^{\td}(w,L)},
\quad
\pp_1(L)=\frac{w f(L)}{Z^{\td}(w,L)},
\quad
\pp_2(L)=\frac{w^2\sum_{ k_1}f( k_1)f(L- k_1)}{Z^{\td}(w,L)},\dots
\ee
and more generally, for $n\ge1$,
\be
\pp_n(L)
=\frac{w^n(f\star)^n(L)}{Z^{\td}(w,L)}=\frac{w^n\left[ \tilde f(z)^n\right]_{L}}{Z^{\td}(w,L)},
\ee
where the notation $[\cdot]_L$ stands for the $L-$th coefficient of the series inside the brackets.
Hence the generating function with respect to $L$ of the numerator of (\ref{eq:pnL}) reads
\beq\label{eq:gfNL}
\sum_{L\ge0}z^L \pp_n(L)\num=w^n\tilde f(z)^n.
\eeq
Taking the sum of the right side upon $n\ge0$ yields back $\tilde Z^{\td}(w,z)$ given in (\ref{eq:fgZ}).

The first moment of this distribution is by definition
\be
\langle N_L\rangle=\sum_{n\ge1}n\, \pp_n(L).
\ee
The generating function of its numerator reads, using (\ref{eq:gfNL})
\beq\label{eq:NLfg}
 \sum_{L\ge0}z^L\mean{N_L}\num=
\sum_{n\ge1}n(w\tilde f(z))^n=\frac{w \tilde f(z)}{\big(1-w \tilde f(z)\big)^2}
=w \frac{\dd \tilde Z^{\td}(w,z)}{\dd w},
\eeq
hence
\beq\label{eq:meanNL}
\mean{N_L}=w\,\frac{\dd \ln Z^{\td}(w,L)}{\dd w},
\eeq
as expected in the grand canonical ensemble with respect to $N_L$.
Alternatively, since
\be
\sum_{L\ge0}z^L\mean{N_L}\num=\tilde Z^{\td}(w,z)^2-\tilde Z^{\td}(w,z),
\ee
we have
\beq\label{eq:niqTdc}
\mean{N_L}\num=(Z^{\td}\star Z^{\td})(w,L)-Z^{\td}(w,L).
\eeq

More generally, the generating function of the moments of $N_L$ is given by
\be
\mean{v^{N_L}}=\sum_{n\ge0}v^n\pp_n(L).
\ee
Taking the generating function with respect to $L$ of the numerator of this expression, using (\ref{eq:gfNL}),
\be
\sum_{L\ge0}z^L\mean{v^{N_L}}\num=\sum_{n\ge0}v^n(w\tilde f(z))^n=\frac{1}{1-vw\tilde f(z)}
=\tilde Z^{\td}(vw,z),
\ee
yields
\beq\label{eq:yNL}
\mean{v^{N_L}}=\frac{Z^{\td}(vw,L)}{Z^{\td}(w,L)}.
\eeq
Likewise, the inverse moment $\mean{1/N_L}$ is
\be
 \Big\langle \frac{1}{N_L}\Big\rangle=\sum_{n\ge1}\frac{\pp_n(L)}{n}
=\frac{1}{Z^{\td}(w,L)}\sum_{n\ge1}\frac{\left[[ w \tilde f(z)]^n\right]_{L}}{n}
=\frac{1}{Z^{\td}(w,L)}\left[-\ln(1-w \tilde f(z))\right]_{L}.
\ee

\subsection{Single interval distribution}
The marginal distribution of one of the summands, say $X_1$,
is by definition
\be
\pi(k|L)=\prob(X_1=k|S_{N_L}=L)=\mean{\delta( X _1,k)},
\ee
where $\mean{\cdot}$ is the average with respect to (\ref{eq:jointe}), with a summation upon the variables $k_1,\dots, k_n$ (with $1\le k\le L$) and $n\ge1$, resulting in
\beqa\label{eq:plN}
 \pi(k|L)\num
&=&\sum_{n\ge1}\sum_{ k_1}\delta( k_1,k) w f( k_1)\sum_{{ k_2,\dots}} w^{n-1}f( k_2)\dots f(k_n)\delta\Big( k_1+\sum_{i=2}^n k_i,L\Big)
\nonumber\\
 &=&\sum_{ k_1}\delta( k_1,k) w f( k_1)\delta( k_1,L)+\sum_{ k_1, k_2}\delta( k_1,k) w^2f( k_1)f( k_2)\delta( k_1+ k_2,L)
+\cdots
\nonumber\\
 &=&w f(k)\delta(k ,L)+w f(k)Z^{\td}(w,L-k)\big(1-\delta(k ,L)\big).
\eeqa
Finally
\beq\label{eq:margin}
 \pi(k|L)
=\stackunder{p(k ,1|L)}{\underbrace{\frac{w f(k)}{Z^{\td}(w,L)}\delta(k ,L)}}+w f(k)\frac{Z^{\td}(w,L-k)}{Z^{\td}(w,L)}\big(1-\delta(k ,L)\big),
\eeq
where the first term corresponds to $n=1$, i.e.,
\beq\label{eq:pLLtdrp}
\pi(L|L)=\frac{w f(L)}{Z^{\td}(w,L)}=\prob(N_L=1).
\eeq
Also, since $Z^{\td}(w,0)=1$, (\ref{eq:margin}) can be more compactly written as
\beq\label{eq:pkL}
\pi(k|L)=\frac{w f(k)Z^{\td}(w,L-k)}{Z^{\td}(w,L)}.
\eeq
The generating function of the numerator of (\ref{eq:pkL}) with respect to $L$ yields
\beq\label{eq:fgplN+}
 \sum_{L\ge k}z^L\pi(k|L)\num=w z^{k} f(k)\tilde Z^{\td}(w,z)
=\frac{w z^{k} f(k)}{1-w \tilde f(z)}.
\eeq
Summing (\ref{eq:pkL}) upon $k$ we obtain 
\beq\label{eq:Zrecur}
Z^{\td}(w,L)=\sum_{k=1}^{L} w f(k)Z^{\td}(w,L-k), \quad L\ge1,
\eeq
which can also be obtained by multiplying the recursion (\ref{eq:recurzrp}) for $Z_n(L)$ by $w^n$ and summing on $n$.

\subsection*{Remarks}
1. An alternative route to (\ref{eq:pkL}) is as follows.
We have
\bea
\pi(k|L)\num&=&\sum_{n\ge1}w^n \sum_{\{k_i\}}\delta(k_1,k)p_n(\{k_i\}|L)\num
=\sum_{n\ge1}w^n \pi_n(k|L)\num
\nonumber\\
&=&\sum_{n\ge1}w^nf(k)Z_{n-1}(L-k)=wf(k)\sum_{n\ge0}w^nZ_n(L-k)
\nonumber\\
&=& wf(k)Z^{\td}(w,L-k)\num.
\eea
2.
We also note that
\be
\pi(k|L)=\sum_{n\ge1}\pi_n(k|L)\pp_n(L),
\ee
which is a simple consequence of (\ref{eq:Newppnpn}).

\subsection{Mean interval $\xmtd$ }
\label{sub:meaninterval}

This is, by definition,
\be
\xmtd=\sum_{k\ge1}k\pi(k|L).
\ee
Multiplying (\ref{eq:fgplN+}) by $k$ and summing upon $k$ yields
\beq\label{eq:fgtaum}
\sum_{L\ge1}z^L
\xmtd \num=\frac{w z\tilde f'(z)}{1-w \tilde f(z)},
\eeq
which can also be obtained by taking the derivative with respect to $z$ of the expression for the inverse moment $\mean{1/N_L}$.
Indeed,
\beq\label{eq:LNL}
\xmtd\num=\Big\langle \frac{L}{N_L}\Big\rangle\num
=L\left[-\ln(1-w \tilde f(z))\right]_{L},
\eeq
then, taking the generating function of the right side gives
\bea
\sum_{L\ge0}z^L\xmtd\num
&=&\sum_{L\ge 1} z^L L\left[-\ln(1-w \tilde f(z))\right]_{L}
\nonumber\\
&=&z\frac{\dd}{\dd z}\left(-\ln(1-w \tilde f(z))\right)=\frac{w z\tilde f'(z)}{1-w \tilde f(z)}.
\eea

\subsection{The longest interval}
\label{sub:longestTDRP}
By definition, the longest interval is
\be
X_{\max}=\max(X_1,\dots,X_{N_L}).
\ee
Its distribution function is defined as
\beq\label{eq:longestTD}
F(k|L)=\prob(X_{\max}\le k|L)
=\sum_{n\ge0}\sum_{k_1=1}^{k}\dots\sum_{k_n=1}^{k}p(\{k_i\},n|L)
=\frac{F(k|L)\num}{Z^{\td}(w,L)},
\eeq
with initial value
\beq\label{eq:F1k0num}
F(k|0)\num=1.
\eeq
The numerator in (\ref{eq:longestTD}) reads
\beqa
F(k|L)\num
&=&\sum_{n\ge0}w^n
\sum_{k_1=1}^{k}\dots\sum_{k_n=1}^{k}p_n(\{k_i\}|L)\num
\nonumber\\
&=&\sum_{n\ge0}w^n F_n(k|L)\num,
\label{eq:F1klnum}
\eeqa
where $F_n(k|L)\num$ is defined in (\ref{eq:numeratorzrp}).
Note that $F(L|L)\num=Z^{\td}(w,L)$, hence $F(L|L)=1$.
The generating function of the numerator is
\bea
\sum_{L\ge0}z^L F(k|L)\num
&=&1+\sum_{n\ge1}\prod_{i=1}^n\Big(\sum_{k_i=1}^{k} wf(k_i)z^{k_i}\Big)
\nonumber\\
&=&1+\sum_{n\ge1}\Big(w\tilde f(z,k)\Big)^n
=\frac{1}{1-w\tilde f(z,k)},
\eea
where
\be
\tilde f(z,k)=\sum_{j=1}^{k} z^{j}f(j).
\ee
The numerator obeys the recursion (renewal) equation, which generalises the Ros\'en-Wendel's result (2.4) of \cite{wendel1},
\be
F(k|L)\num=\sum_{j=1}^{\min(k ,L)}wf(j)\,F(k|L-j)\num,
\ee
with initial condition (\ref{eq:F1k0num}).

The distribution of $X_\max$ is given by the difference
\be
p(k|L)=\prob(X_{\max}=k)=F(k|L)-F(k -1|L),
\ee
where $F(0|L)=\delta(L,0)$,
with generating function
\beqa\label{eq:p1gf}
\sum_{L\ge0}z^Lp(k|L)\num
&=&\frac{1}{1-w\tilde f(z,k)}-\frac{1}{1-w\tilde f(z,k-1)}
\nonumber\\
&=&\frac{wz^{k}f(k)}{[1-w\tilde f(z,k)][1-w\tilde f(z,k-1)]}.
\eeqa
Its end point value is the same as $\pi(L|L)$ (\ref{eq:pLLtdrp}), i.e.,
\beq\label{eq:terminal}
p(L|L)=\prob(N_L=1)=\frac{wf(L)}{Z^{\td}(w,L)}.
\eeq

Note that (\ref{eq:p1gf}) can be obtained by multiplying (\ref{eq:p1gen}) by $w^n$ and summing on $n$.
In other words,
\beq\label{eq:new1}
p(k|L)\num=\sum_{n\ge1}w^np_n(k|L)\num,
\eeq
as can also be inferred from (\ref{eq:F1klnum}).
And therefore (see (\ref{eq:Newppnpn}))
\beq\label{eq:new1+}
p(k|L)=\sum_{n\ge1}p_n(k|L)\pp_n(L).
\eeq

When $X_\max=k>L/2$, the longest interval is unique.
Denoting the restriction of $p(k|L)$ to the range $k>L/2$ by $ q(k|L)$,
and generalising the reasoning made in \cite{wendel,wendel1} we can decompose a configuration into three contributions to obtain
\beqa\label{eq:qkL}
q(k|L)\num&=&\sum_{i=0}^{L-k}Z^{\td}(w,i)wf(k)Z^{\td}(w,L-k-i)
\nonumber\\
&=&wf(k)(Z^{\td}\star Z^{\td})(w,L-k),
\eeqa
which, using (\ref{eq:niqTdc}), can be alternatively written as
\beq\label{eq:tdc}
 q(k|L)
=\frac{wf(k)Z^{\td}(w,L-k)}{Z^{\td}(w,L)}(1+\mean{N_{L-k}}),
\eeq
that is (see (\ref{eq:pkL})),
\beq\label{eq:better}
 q(k|L)
=\pi(k|L)(1+\mean{N_{L-k}}).
\eeq
For $k=L$, (\ref {eq:terminal}) with (\ref{eq:pLLtdrp}) is recovered.

 \subsection*{Remarks} 

1. An alternative route to 
(\ref{eq:better}) can be inferred from (\ref{eq:new1}) as follows.
We have
\beq\label{eq:variant}
 q(k|L)\num=\sum_{n\ge1}w^nq_n(k|L)\num=\sum_{n\ge1}w^n n \pi_n(k|L)\num,
 \eeq
 where $\pi_n(k|L)$ is given in (\ref{eq:zrppkL}).
 So
 \beqa\label{eq:variante}
 q(k|L)\num
&=&wf(k)\sum_{n\ge1}nw^{n-1} Z_{n-1}(L-k)
 \nonumber\\
&=&wf(k)\sum_{n\ge0}(w^nZ_n(L-k)+nw^n Z_n(L-k))
\nonumber\\
&=&wf(k)(Z^{\td}(w,L-k)+\mean{N_{L-k}}\num),
 \eeqa
 which, after division by $Z^{\td}(w,L)$, yields (\ref{eq:better}).
 
 \smallskip
 \noindent 2.
 Comparison between (\ref{eq:variante}) and (\ref{eq:qkL}) entails 
the equality
 \be
(Z^{\td}\star Z^{\td})(w,L)=\sum_{n\ge1}nw^{n-1}Z_{n-1}(L),
\ee
which can also be checked directly by taking the generating functions of both sides,
\be
\tilde Z^{\td}(w,z)^2=\sum_{n\ge1}n w^{n-1}\tilde f(z)^{n-1}=\frac{1}{(1-\tilde wf(z))^2}.
\ee
 
 \smallskip
 \noindent 3.
 From (\ref{eq:FnZn}) we infer that, if $k>L/2$, 
 \beq\label{eq:FtdZtd}
 F^{\td}(k|L-k)\num=Z^{\td}(w,L-k).
 \eeq

\section{Phase transition for tied-down renewal processes}

\label{sec:regimetdrp}

In this section and in the following sections \ref{sec:tdrp-disordered}, \ref{sec:tdrp-critical} and \ref{sec:tdrpcondensed}
the distribution $f(k)=\prob(X=k)$ is taken subexponential with asymptotic power-law decay (\ref{eq:powerlaw}).
Before discussing the phase diagram of the process, we give some illustrative examples of such distributions.

\subsection{Illustrative examples}

In the sequel, we shall illustrate the general results derived for \textsc{tdrp} in the current section, and for free renewal processes in section \ref{sec:rpGeneral}, on the following examples.

\medskip\noindent
\textit{Example 1.} 
This first example corresponds to the tied-down random walk of figure \ref{fig:tdrw}
on which we come back in more detail.
The distribution of the size of intervals, $f(k)$, representing the probability of first return at the origin of the walk after $2k$ steps, or equivalently after $k$ tick marks on figure \ref{fig:tdrw}, reads
\beq\label{eq:ex1td}
f(k)=\frac{1}{2^{2 k-1}}\frac{(2 k-2)!}{( k-1)! k!}=\frac{\Gamma( k-1/2)}{2\sqrt{\pi}\Gamma( k+1)}
\approx \frac{1}{2\sqrt{\pi}k^{3/2}},
\eeq
since the number of such walks is equal to $(2 k-2)!/[( k-1)! k!]$.
Its generating function reads
\be
\tilde f(z)=1-\sqrt{1-z}.
\ee
The partition function (\ref{eq:Z1L}) for $w=1$
represents the probability that the walk returns at the origin after $2L$ steps, or equivalently after $L$ tick marks,
\beq\label{eq:Z1Ltdrw}
Z^{\td}(1,L)=\frac{1}{2^{2L}}\bino{2L}{L}
\approx \frac{1}{\sqrt{\pi L}},
\eeq
since the number of such walks is equal to $(2 L)!/(L!)^2$.
Its generating function reads
\be
\tilde Z^{\td}(1,z)=\frac{1}{\sqrt{1-z}}.
\ee
Note that
\be
f(L)=Z^{\td}(1,L-1)-Z^{\td}(1,L).
\ee
The partition function $Z_n(L)$ is explicit for this case, 
\beq\label{eq:ZnL-toy}
Z_n(L)=\frac{n}{2^{2 L - n}}\frac{(2 L - n - 1)!}{L! (L - n)!}
 \approx \frac{n}{2\sqrt{\pi}L^{3/2}},
\eeq
with $n\le L$.

\medskip\noindent
\textit{Example 2.}
This second example is defined for any $\th>0$ by
\beq\label{eq:fkex2}
f(k)=\frac{1}{\zeta(1+\th)}\frac{1}{k^{1+\th}},\quad k>0.
\eeq
So
\be
\tilde f(z)=\frac{\mathrm{Li}_{1+\th}(z)}{\zeta(1+\th)},
\ee
where $\mathrm{Li}_{s}(z)=\sum_{k\ge1}z^k/k^s$ is the polylogarithm function.
If $\th>1$ the mean $\xm=\zeta(\th)/\zeta(1+\th)$.
This is the distribution used, e.g., in \cite{burda,burda3,bar2,bar3,barma,glzeta}.

\subsection{Phase diagram}
Demonstrating the existence of a phase transition in the model defined by (\ref{eq:jointe}), with distribution 
$f(k)$ given by (\ref{eq:powerlaw}), when $w$ crosses the value one, is a classical subject.
This model is a particular instance of a linear system, as described in \cite{fisher}, where the mechanism of the transition is explained in simple terms.
This transition is also studied in \cite{burda3,bar2,bar3,barma} for \textit{Example 2} (see (\ref{eq:fkex2})).
Let us first analyse the large $L$ behaviour of $Z^{\td}(w,L)$.
Recalling (\ref{eq:fgZ})
we have, for a contour encircling the origin,
\be
Z^{\td}(w,L)=\oint\frac{\dd z}{2\pi\ii }\frac{\tilde Z^{\td}(w,z)}{z^{L+1}}=
\oint\frac{\dd z}{2\pi\ii \,z^{L+1}}\frac{1}{1-w\tilde f(z)}.
\ee
Since $\tilde f(z)$ is monotonically increasing for $z\in(0,1)$ the denominator of $\tilde Z^{\td}(w,z)$, $1-w\tilde f(z)$, is monotonically decreasing between 1 and $1-w$.

\smallskip\noindent \textbf{Disordered phase.} 
If $w>1$, the denominator vanishes for $z=z_0<1$ such that $w\tilde f(z_0)=1$, hence $\tilde Z^{\td}(w,z)$ has a pole at $z_0$, and therefore $Z^{\td}(w,L)$ is exponentially increasing,
\beq\label{eq:Zaswgt1}
Z^{\td}(w,L)\approx \frac{z_0^{-L}}{wz_0\tilde f'(z_0)}.
\eeq

\smallskip \noindent \textbf{Critical regime.} 
If $w=1$, then $z_0=1$.
The asymptotic estimates of $Z^{\td}(1,L)$ are given in (\ref{eq:Z1asympt}) and (\ref{eq:Ztdgt1}).

\smallskip \noindent \textbf{Condensed phase.} 
If $w<1$, the denominator $1-w\tilde f(z)$ has no zero, but it is singular for $z=z_0=1$ (which is the singularity of 
$\tilde f(z)$).
Hence $z_0$ sticks to 1.
The asymptotic estimate of $Z^{\td}(w,L)$ is given in (\ref{eq:ZLwlt1}).

This is the \textit{switch} mechanism of Fisher \cite{fisher}: the condition determining $z_0$ switches from $z_0$ being the smallest root of the equation $1-w\tilde f(z)=0$ to being the closest real singularity of $\tilde f(z)$, which is a cut at $z=z_0=1$.
This non analytical switch signals the phase transition.
The free energy density \cite{fisher}
\be
\mathbf{f}=\lim_{L\to\infty}-\frac{1}{L}\ln Z^{\td}(w,L)=\ln z_0,
\ee
therefore vanishes when $w\le1$.
The three cases above are successively reviewed in the next sections.

\section{Disordered phase ($w>1$) for tied-down renewal processes}
\label{sec:tdrp-disordered}

The asymptotic expression at large $L$ of the distribution of the size of a generic interval is obtained by
carrying (\ref{eq:Zaswgt1}) in (\ref{eq:pkL}), which leads to
\beq\label{eq:pkLsuper}
\pi(k|L)\approx wf(k)z_0^k=wf(k)\e^{-k/\xi},\qquad \xi=\frac{1}{|\ln z_0|},
\eeq
where $\xi$ is the correlation length, divergent at the transition.
This expression is independent of $L$ and normalised, since summing on $k$ restores $w \tilde f(z_0)=1$.
This exponentially decaying distribution has a finite mean,
\beq\label{eq:XLwgt1}
\xmtd \approx wz_0\tilde f'(z_0),
\eeq
an expression which can also be inferred from (\ref{eq:fgtaum}).
Thus (\ref{eq:Zaswgt1}) can be recast as
\be
Z^{\td}(w,L)\approx \frac{z_0^{-L}}{\xmtd}.
\ee
The distribution of $N_L$ is given by (\ref{eq:pnL})
\be
\pp_n(L)=\frac{w^nZ_n(L)}{Z^{\td}(w,L)}\approx w^nZ_n(L)\, w\tilde f'(z_0)z_0^{L+1}.
\ee
This distribution obeys the central limit theorem, as illustrated on the example below.
Using (\ref {eq:meanNL}), we obtain the asymptotic expression of $\mean{N_L}$,
\be
\mean{N_L}\approx -L\frac{w}{z_0}\frac{\dd z_0}{\dd w}\approx \frac{L}{\xmtd},
\ee
which means that
\be
\frac{1}{\mean{N_L}}\approx \bigmean{\frac{1}{N_L}}.
\ee
Let us denote the density of points (or intervals) for a finite system as
\beq\label{eq:nuL}
\nu_L=\frac{\mean{N_L}}{L},
\eeq
then, asymptotically, we have 
\beq\label{eq:defnu}
\nu=\lim_{L\to\infty} \nu_L=\lim_{L\to\infty}\frac{\mean{N_L}}{L}=\lim_{L\to\infty}\frac{1}{\xmtd}.
\eeq

\begin{figure}[!ht]
\begin{center}
\includegraphics[angle=0,width=.8\linewidth]{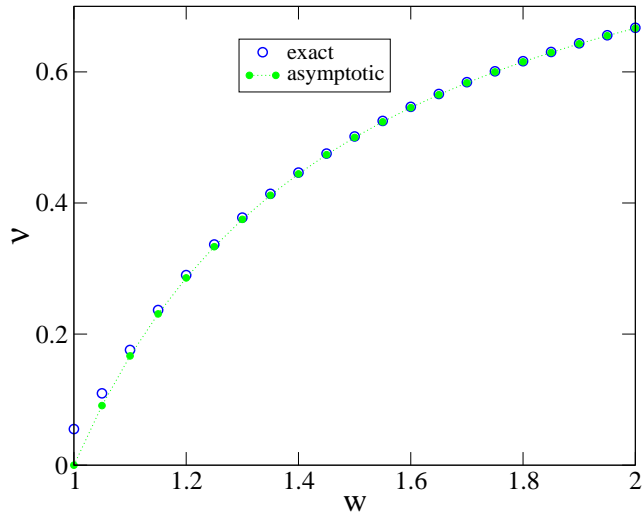}
\caption{\textsc{tdrp}: density of intervals $\nu$ against $w$ in the disordered phase, for \textit{Example 1} (see (\ref{eq:ex1td})). 
\textit{Exact} refers to the finite-size expression (\ref{eq:nuL}) where $\mean{N_L}$ is extracted from (\ref{eq:NLfg}) with $L=1000$, \textit{asymptotic} to (\ref{eq:rho}).
}
\label{fig:density}
\end{center}
\end{figure}

We illustrate these general statements on \textit{Example 1} (see (\ref{eq:ex1td})),
for which $z_0$ is explicit,
\be
z_0=\frac{2w-1}{w^2},
\ee
hence
\be
\xi\approx (w-1)^{-2},
\ee
and the following asymptotic expressions hold,
\beqa
Z^{\td}(w,L)&\approx& \frac{2(w-1)w^{2L}}{(2w-1)^{L+1}},
\nonumber\\\xmtd&\stackunder{L\to\infty}{\to}& \frac{2w-1}{2(w-1)},
\nonumber\\\mean{N_L}&\approx& \frac{L}{\xmtd}+\frac{w}{(w-1)(2w-1)},
\nonumber\\\var N_L&\approx& L\frac{2w}{(2w-1)^2}-\frac{w(2w^2-1)}{(2w-1)^2(w-1)^2},
\nonumber\\ \pp_n(L)&\approx&\frac{1}{\sqrt{2\pi\var{N_L}}}\exp\Big(-\frac{(n-\mean{N_L})^2}{2\var{N_L}}\Big),
\nonumber\\ \nu&=&\frac{2(w-1)}{2w-1}.
\label{eq:rho}
\eeqa
Figure \ref{fig:density} depicts a comparison between the exact finite-size expression of the density $\nu_L$ obtained by means of (\ref{eq:NLfg}) for $L=1000$ as a function of $w$, and the asymptotic expression (\ref{eq:rho}).
It vanishes at the transition $w=1$, where the system becomes critical.

More generally, if $\th<1$, close to the transition, we get
\be
\nu\sim(w-1)^{1/\th-1},
\ee
as can be easily inferred by means of the expansion (\ref{eq:expansionLaplace}), a result already present in \cite{burda3}, later recovered in \cite{bar2}.
If $\th>1$, the density $\nu$ tends to $1/\xm$ when $w\to1$, as can be seen on (\ref{eq:XLwgt1}) and (\ref{eq:defnu}), using the expansion (\ref{eq:expansionLaplacegt1}) (see also (\ref{eq:Njuste})).
The density vanishes in the condensed phase since $\mean{N_L}$ is finite (see (\ref{eq:NLTDcond})), it is therefore discontinuous at the transition, as noted in \cite{burda3,bar2}.
Likewise, it is easy to see that
\beq\label{eq:xi}
\xi^{-1}\sim \left\{
\begin{array}{ll}
(w-1)^{1/\th}& \th<1
\vspace{12pt}
\\
w-1& \th>1.
\end{array}
\right.
\eeq
The correlation length diverges at the transition, while the order parameter $\nu$ is either continuous ($\th<1$) or discontinuous ($\th>1$), as seen above.
The transition is therefore of mixed order \cite{burda3,bar2}. 

Finally we note that the intervals $X_i$ behave essentially as iid random variables, with distribution (\ref{eq:pkLsuper}), hence the statistics of the longest interval belongs to the Gumbel class \cite{gumbel,gnedenko}.
This is detailed on \textit{Example 2} in \cite{bar3}. \\

\section{Critical regime ($w=1$) for tied-down renewal processes}
\label{sec:tdrp-critical}

In this regime, the behaviour of the quantities of interest strongly depends on whether the index $\th$ is smaller or larger than unity.
The discussion below is organised accordingly.
Part of the material of this section can be found in more details in \cite{wendel,wendel1} and is also addressed in \cite{bar3,barma}.
Here we summarise these former studies and complement them by a detailed analysis of the distribution of the number of intervals $N_L$ and of the distribution $\pi(k|L)$ of the size of a generic interval.
We also come back on the distribution of the longest interval.
 
If $w=1$ the singularity is at $z=1$, or, setting $z=\e^{-s}$, at $s=0$.
The generating function $\tilde f(z)$ becomes the Laplace transform $\hat f(s)$
which has the expansion
\beqa\label{eq:expansionLaplace}
\hat f(s)\stackunder{s\to0}{\approx} 1-|a| s^\th, \qquad &\th<1
\\
\label{eq:expansionLaplacegt1}
\hat f(s)\stackunder{s\to0}{\approx} 1-s\xm +\cdots+a s^\th,\qquad &\th>1
\eeqa
with
\be
a=c\,\Gamma(-\th)=\th\Gamma(-\th)k_0^\th,
\ee
i.e., $c=\th k_0^\th$, where $k_0$ is a microscopic scale, defined as
\beq\label{eq:defg}
 g(k)=\sum_{j> k} f(j)\stackunder{k\to\infty}{\approx} \Big(\frac{k_0}{k}\Big)^\th.
\eeq
The parameter $a$ is negative if $0<\theta<1$, positive if $1<\theta<2$, and so on.
For instance, $\Gamma(-1/2)=-2\sqrt{\pi}$, $\Gamma(-3/2)=4\sqrt{\pi}/3$, $\Gamma(-5/2)=-8\sqrt{\pi}/15$, and so on.

\subsection{Distribution $f(k)$ with index $\th<1$}

Since $\tilde Z^{\td}(1,z)=1/(1-\tilde f(z))$, in Laplace space we have $\hat Z^{\td}(1,s)\approx 1/as^{\th}$
which yields the expression of the partition function (see (4.6) in \cite{wendel1}) 
\beq\label{eq:Z1asympt}
Z^{\td}(1,L)\stackunder{L\to\infty}{\approx} \frac{\th\sin\pi\th}{\pi c}L^{\th-1}.
\eeq
For instance, setting $\th=1/2$ and $c=1/(2\sqrt{\pi})$ restores (\ref{eq:Z1Ltdrw}).

\subsubsection{The number of intervals}
We have (see (4.10) in \cite{wendel1}),
\beq\label{eq:NLaveragecrit}
\mean{N_L}\approx \frac{A(\th)}{c}L^\th,\qquad A(\th)=\frac{\Gamma(1+\th)}{\Gamma(1-\th)\Gamma(2\th)},
\eeq
which can be easily deduced from (\ref{eq:NLfg}).
For the specific case of \textit{Example 1} (see (\ref{eq:ex1td})),
we have the exact result (see (2.47) in \cite{wendel1})
\be
\langle N_L\rangle=\frac{1}{Z^{\td}(1,L)}-1=\frac{2^{2L}}{\bino{2L}{L}}-1\approx\sqrt{\pi L},
\ee
which is in agreement with (\ref{eq:NLaveragecrit}), with $\th=1/2,c=1/(2\sqrt{\pi})$.
We know from (\ref{eq:pnL}) that the distribution of $N_L$ is given by the ratio
\be
\pp_n(L)=\frac{Z_n(L)}{Z^{\td}(1,L)}.
\ee
For $n$ and $L$ large $\pp_n(L)$ has a scaling form.
On the one hand, according to the generalised central limit theorem, the scaling form of the numerator is given by
\be
Z_n(L)\approx \frac{1}{n^{1/\th}}\mathcal{L}_{\th,c}\bigac{\frac{L}{n^{1/\th}}},
\ee
where $\mathcal{L}_{\th,c}$ is the density of the stable law of index $\th$, tail parameter $c$ and asymmetry parameter $\beta=1$ (see e.g., \cite{cg2019}).
Then using (\ref{eq:Z1asympt}), we get, with $u=L/n^{1/\th}$,
\be
\pp_n(L)\approx \frac{\pi c}{\th\sin\pi\th}\frac{1}{L^\th}\frac{L}{n^{1/\th}}\mathcal{L}_{\th,c}\bigac{\frac{L}{n^{1/\th}}}
\approx \frac{\pi c}{\th\sin\pi\th}\frac{1}{L^\th}u\mathcal{L}_{\th,c}(u).
\ee
The L\'evy distribution of index $\th=1/2$ has the explicit expression
\beq\label{eq:stable12Lev}
\mathcal{L}_{1/2,c}(u)=\frac{c\,\e^{-\pi c^2/u}}{u^{3/2}},
\eeq
hence (see (2.49) in \cite{wendel1}), for \textit{Example 1} (see (\ref{eq:ex1td})),
\be
\pp_n(L)\approx \frac{v}{2\sqrt{L}}\e^{-v^2/4}, \quad v=\frac{1}{\sqrt{u}}=\frac{n}{\sqrt{L}}.
\ee
Moreover, for this example, for $n$ and $L$ finite, $\pp_n(L)$ is explicit since both $Z_n(L)$ given by (\ref{eq:ZnL-toy}) and $Z^{\td}(1,L)$, given by (\ref{eq:Z1Ltdrw}), are known explicitly.

\begin{figure}[!ht]
\begin{center}
\includegraphics[angle=0,width=.8\linewidth]{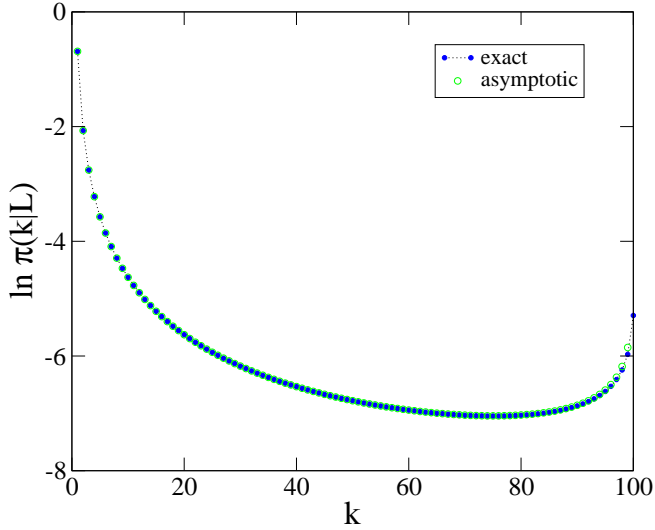}
\caption
{\textsc{tdrp}: single interval distribution $\pi(k|L)$, with $L=100$, for
\textit{Example 1} (see (\ref{eq:ex1td})), at criticality ($w=1$).
\textit{Exact} refers to the middle expression in (\ref{eq:pkLestim}), \textit{asymptotic} to the rightmost one.
The $y-$axis is on a logarithmic scale.}
\label{fig:pkL}
\end{center}
\end{figure}

%
\subsubsection{Single interval distribution}
Two regimes are to be considered.

\noindent {\bf (i)}
In all regimes where $\l=L-k$ is large,
using (\ref{eq:Z1asympt}) we have
\beq\label{eq:pkLestim}
\pi(k|L)= f(k)\frac{Z^{\td}(1,L-k)}{Z^{\td}(1,L)}\approx f(k)\left(1-\frac{k}{L}\right)^{\th-1}.
\eeq
For instance, if $1\ll k\ll L$,
\be
\pi(k|L)\approx f(k)\approx \frac{c}{k^{1+\th}},
\ee
while if $k=\la L$, with $\la\in(0,1)$,
\be
\pi(k|L)\approx\frac{c}{\la^{1+\th}(1-\la)^{1-\th}}\frac{1}{L^{1+\th}},
\ee
which is minimum at $\la=(1+\th)/2$.

\noindent {\bf (ii)} On the other hand, if $\l=L-k=O(1)$,
\be
\pi(L-\l|L)=f(L-\l)\frac{Z^{\td}(1,\l)}{Z^{\td}(1,L)}\approx \frac{\pi c^2 Z^{\td}(1,\l)}{\th\sin\pi\th }\frac{1}{L^{2\th}}.
\ee
In particular, for $k=L$, 
\be
\pi(L|L)=\frac{f(L)}{Z^{\td}(1,L)}\approx \frac{\pi c^2}{\th\sin\pi\th}\frac{1}{L^{2\th}}.
\ee
In this regime the ratio of $\pi(k|L)$ to the estimate (\ref{eq:pkLestim}) reads
\be
\frac{\pi(L-\l|L)}{f(L-\l)(1-k/L)^{\th-1}}
=\frac{\pi c\,\l^{1-\th}Z^{\td}(1,\l)}{\th\sin\pi\th},
\ee
which tends to one when $\l$ becomes large, if one refers to (\ref{eq:Z1asympt}).

All these results can be illustrated on \textit{Example 1} (see (\ref{eq:ex1td})).
For instance figure \ref{fig:pkL} gives a comparison between the exact expression of $\pi(k|L)$ computed for $L=100$ by means of the middle expression in (\ref{eq:pkLestim}) and its asymptotic form given by the rightmost expression in (\ref{eq:pkLestim}).

The generating function of the mean interval $\xmtd$ given in (\ref{eq:fgtaum}) yields the estimate, in Laplace space,
\be
\sum_{L\ge1}z^L
\xmtd \num
\approx-\frac{\hat f'(s)}{1-\hat f(s)}\approx \frac{\th}{s},
\ee
hence (see (4.18) in \cite{wendel1})
\beq\label{eq:xmtd-crit}
\xmtd\approx\frac{\th}{Z^{\td}(1,L)}\approx \frac{\pi c}{\sin \pi\th} L^{1-\th}.
\eeq
This result can be recovered by taking the average of the estimate (\ref{eq:pkLestim}).
It predicts correctly that the product $\xmtd \mean{N_L}\sim L$.
For \textit{Example 1}, (see (\ref{eq:ex1td})), the computation leads to the exact result (see (2.58) in \cite{wendel1})
\be
\xmtd=\frac{1}{2Z^{\td}(1,L)}\approx \frac{\sqrt{\pi L}}{2}.
\ee

\subsubsection{The longest interval}
The study of the statistics of the longest interval for the critical case, including the scaling analysis of $p(k|L)$ for $\th<1$, is done in \cite{wendel,wendel1} (see also \cite{bar3} for a study of \textit{Example 2}).

For $k>L/2$, in contrast with the case of random allocation models and \textsc{zrp} where $p(k|L)=q(k|L)=n \pi(k|L)$ (see \S \ref {sec:zrpLargest}), the `enhancement factor' $n$ is now replaced by the factor $1+\mean{N_{L-k}}$ (see (\ref{eq:better})), which is equal to one for $k=L$, where $p(L|L)=\pi(L|L)$ are equal, see (\ref{eq:terminal}).
Using (\ref{eq:NLaveragecrit}) for $\mean{N_L}$, and (\ref{eq:pkLestim}) for $\pi(k|L)$ in (\ref{eq:better})
allows to recover the universal scaling expression valid for $k>L/2$ (see equation (4.48) in \cite{wendel1}), in the limit where $1\ll k\sim L$, with $r=k/L$ kept fixed,
\be
q(k|L)\approx \frac{1}{L}\frac{A(\th)}{r^{1+\th}(1-r)^{1-2\th}},
\ee
where $A(\th)$ is given in (\ref{eq:NLaveragecrit}).

Though there is no condensation at criticality, some features are precursors of this phenomenon.
For instance, the mean longest interval $\mean{X_\max}$ scales as $L$ while the typical interval $\xmtd$ scales as $L^{1-\th}$.
However, not only $X_\max\equiv X^{(1)}$ scales as $L$ but also all
the following maxima $X^{(r)}$ ($k=2,3,\dots$) do so \cite{wendel,wendel1}.
Moreover $X_\max$ continues to fluctuate when $L\to\infty$ while for genuine condensation as in section \ref {sec:zrp} above or in section \ref{sec:tdrpcondensed} below, its distribution is peaked.
Finally, the dominant contribution to the weight of $\pi(k|L)$ comes from values of $k$ less than a small cutoff.

\subsection{Distribution $f(k)$ with index $\th>1$}
For $\th>1$, we have, using (\ref{eq:fgZ}) (see (4.73) in \cite{wendel1}), 
\beq\label{eq:Ztdgt1}
Z^{\td}(1,L)\approx \frac{1}{\xm}+\frac{c}{\th(\th-1)\mean{X}^2}L^{1-\th}.
\eeq
The average value of $N_L$ is obtained by means of (\ref{eq:NLfg})\footnote{Equation (\ref{eq:Njuste}) corrects the inaccurate expression (4.74) given in \cite{wendel1} for this quantity.}
\beq\label{eq:Njuste}
\mean{N_L}\approx\left\{
\begin{array}{ll}
\frac{L}{\xm}+\frac{c}{(\th-1)(2-\th)\mean{X}^2} L^{2-\th}& 1<\th<2
\vspace{12pt}
\\ 
\frac{L}{\xm}+\frac{\var{X}}{\xm^2} & \th>2
\end{array}
\right.
\eeq

The subleading correction in the second line (i.e., for $\th>2$) is given by the correction term of the first line which is now negative and decreasing.
The distribution of $N_L$ reads
\be
\pp_n(L)=\frac{Z_n(L)}{Z^{\td}(1,L)}\approx \xm Z_n(L).
\ee
The asymptotic estimate for $\xmtd$ is obtained by analysing (\ref{eq:fgtaum}), yielding for $\th>1$,
\be
\xmtd\approx\xm-\frac{c}{\th-1}L^{1-\th},
\ee
which is the same expression as for a free renewal process \cite{gl2001}. 
The single interval distribution has the form 
\be
\pi(k|L)= f(k)\frac{Z^{\td}(1,L-k)}{Z^{\td}(1,L)}
\stackunder{L\to\infty}{\approx} f(k),
\ee
except for $L-k$ finite, where in particular,
\be
\pi(L|L)=\frac{f(L)}{Z^{\td}(1,L)}\approx f(L)\xm.
\ee

The distribution of the longest interval is analysed in \cite{wendel1} (see also \cite{bar3} for the case of \textit{Example 2} defined in (\ref{eq:fkex2})).
The result is
\be
F(k|L)\approx \e^{-[L/\xm](k_0/k)^{\theta}},
\ee
where $k_0$ is related to the tail coefficient by $c=\th k_0^\th$.
Setting
\be
X_\max=k_0\left(\frac{L}{\xm}\right)^{1/\theta}Y_L,
\ee
we have, as $L\to\infty$, $Y_L\to Y^{F}$, with limiting distribution 
\be\label{eq:frechet}
\prob(Y^{F}<x)=\e^{-1/x^\theta},
\ee
which is the Fr\'echet law \cite{frechet,gnedenko}.
Therefore
\be
\langle X_{\max}\rangle\approx k_0\left(\frac{L}{\xm}\right)^{1/\theta}\underbrace{\langle Y^{F}\rangle}_{\Gamma(1-1/\theta)},
\ee
as was already the case for free renewal processes \cite{gms2015}.

\section{Condensed phase ($w<1$) for tied-down renewal processes}
\label{sec:tdrpcondensed}

The aim of this section---central in the present work---is to investigate the statistics of the number of intervals and characterise the fluctuations of the condensate.
We start by analysing the large $L$ behaviour of the quantities of interest which are functions of $L$ only (partition function, moments and distribution of $N_L$).
We then investigate the regimes for the distributions of the size of a generic interval, $\pi(k|L)$, and of the longest one, $p(k|L)$.
Related material can be found in \cite{gia1,bar3,barma}.
\subsection{Asymptotic estimates at large $L$}

Starting from (\ref{eq:fgZ}) and linearising with respect to the singular part, we obtain, when $L\to\infty$,
for any value of $\th$,
\beq\label {eq:ZLwlt1}
Z^{\td}(w,L)\approx \frac{w }{(1-w)^2}\frac{c}{L^{1+\th}}\approx \frac{w }{(1-w)^2}f(L).
\eeq
Alternatively, it suffices to notice that, for $n$ fixed and $L$ large, for any subexponential distribution \cite{chistyakov}, 
\beq\label{eq:nfixed}
Z_n(L)\approx n f(L),
\eeq
as for example in (\ref{eq:ZnL-toy}), which entails 
\be
Z^{\td}(w,L)=\sum_{n\ge0}w^nZ_n(L)\approx \sum_{n\ge0}w^n n f(L),
\ee
hence restores (\ref{eq:ZLwlt1}). 
Likewise, we find
\beq\label{eq:Z2}
(Z^{\td}\star Z^{\td})(w,L)\approx \frac{2w }{(1-w)^3}f(L).
\eeq

If one substitutes (\ref {eq:ZLwlt1}) in the expression of the mean $\mean{N_L}$ (\ref{eq:meanNL}), we obtain, the superuniversal result, independent of $\th$,
\beq\label{eq:NLTDcond}
\langle N_L\rangle\stackunder{L\to\infty}{\to}
\frac{1+w }{1-w }.
\eeq
This result can be found alternatively using (\ref{eq:niqTdc}) together with (\ref{eq:ZLwlt1}) and (\ref{eq:Z2}),
or else using (\ref{eq:pnwlt1}) below.
\begin{figure}[!ht]
\begin{center}
\includegraphics[angle=0,width=0.9\linewidth]{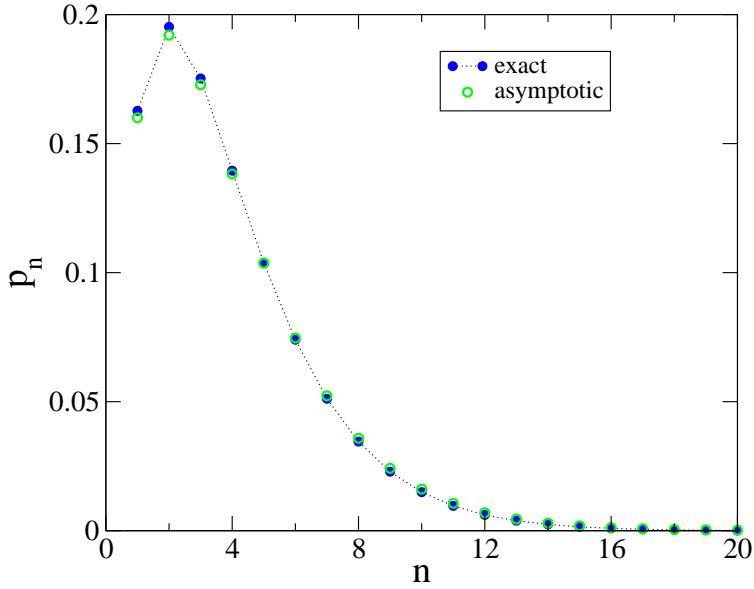}
\caption
{\textsc{tdrp}: superuniversal asymptotic distribution $\pp_n$ of the number $N_L$ of intervals in the condensed phase for $w =0.6$.
\textit{Exact} refers to $\pp_n(L)$ extracted from (\ref{eq:gfNL}) or (\ref{eq:yNL}) for $L=200$ with \textit{Example 1} (see (\ref{eq:ex1td})), \textit{asymptotic} refers to (\ref{eq:pnwlt1}).
For this value of $w$, the asymptotic average $\mean{N_L}=4$.
}
\label{fig:pn}
\end{center}
\end{figure}
More generally, this superuniversality also holds for the asymptotic distribution of $N_L$.
Using (\ref{eq:yNL}), the latter reads
\be
\mean{v^{N_L}}=\sum_{n\ge0}v^n\pp_n(L)=\frac{Z^{\td}(vw,L)}{Z^{\td}(w,L)}\approx \frac{y(1-w)^2}{(1-vw)^2},
\ee
hence extracting the coefficient of order $n$ in $y$ of this expression leads to the asymptotic distribution, independent of $\th$,
\beq\label{eq:pnwlt1}
\pp_n(L)\stackunder{L\to\infty}{\to} \pp_n=n(1-w)^2 w^{n-1}.
\eeq
This distribution is depicted in figure \ref{fig:pn}.
The same result can be found by noting that
\be
\pp_n(L)=\frac{w^n Z_n(L)}{Z^{\td}(w,L)}\approx \frac{Z_n(L)}{f(L)}(1-w)^2 w^{n-1}\stackunder{L\to\infty}{\to} n(1-w)^2 w^{n-1},
\ee
using (\ref{eq:nfixed}) again.
The interpretation of (\ref{eq:pnwlt1}) is simple:
$N_L-1$ is the sum of two independent geometric random variables (see (\ref{eq:geo-free})), which represent the fluid on either side of the remaining interval, which is the condensate.

The inverse moment $\mean{1/N_L}$
can be obtained by using (\ref{eq:pnwlt1}) above, 
\beq\label{eq:unsurN}
\Big\langle \frac{1}{N_L}\Big\rangle\to 1-w.
\eeq
As a consequence of (\ref{eq:unsurN}) we have, for any value of $\th$,
\beq\label{eq:x-average}
\xmtd\approx (1-w)L,
\eeq
which can also be deduced from the asymptotic analysis of (\ref{eq:fgtaum}).

\subsection{Regimes for the single interval distribution}
\label{sub:tdrpregimes}
For $L$ large, the asymptotic expression of the single interval distribution is obtained by
substituting (\ref{eq:ZLwlt1}) in (\ref{eq:margin}), leading to 
\beq\label{eq:margasym}
\pi(k|L)
\stackunder{L\to\infty}{\approx} (1-w)^2 f(k)\frac{Z^{\td}(w,L-k)}{f(L)}.
\eeq
Figure \ref{fig:pp1} depicts the distribution $\pi(k|L)$ (together with the distribution of the longest interval 
$p(k|L)$, see \S \ref{sub:regimesMax} below), for $L=60$ and $w=0.6$ computed from (\ref{eq:pkL}), with \textit{Example 1} (see (\ref{eq:ex1td})).
As can be seen on this figure, there are three distinct regions for $\pi(k|L)$, namely, from left to right,
a downhill region, followed by a long dip region, then an uphill region which accounts for the fluctuations of the condensate.

Let us discuss the behaviour of $\pi(k|L)$ given by (\ref{eq:margasym}) in each of these regions successively.
\begin{enumerate}
\item \textit{Downhill region.} For $k$ finite
we have, using again (\ref{eq:ZLwlt1}),
\beq\label{eq:estimpkL}
\pi(k|L)\approx wf(k).
\eeq
Introducing a cutoff $\Lambda$, such that $1\ll\Lambda\ll L$,
the weight of this downhill region can be estimated as
\beq\label{eq:region1}
\sum_{k=1}^\Lambda \pi(k|L)
\approx \sum_{k=1}^{\Lambda}wf(k)\to\sum_{k=1}^{\infty} wf(k) = w.
\eeq
\item
\smallskip\noindent \textit{Dip region.}
In the dip region, where $k$ and $L-k$ are simultaneously large, setting $k=\la L$ in (\ref{eq:estimpkL}) 
($0<\la<1$) and using (\ref{eq:ZLwlt1}) yields the estimate
\beq\label{eq:dip}
\pi(k|L)\approx w f(k)\frac{f(L-k)}{f(L)}\approx \frac{w}{[\la(1-\la)]^{1+\th}}\frac{c}{L^{1+\th}}.
\eeq
In this region the distribution is therefore U-shaped: the most probable configurations are those where almost all the particles are located on one of two sites. 
The dip centred around $k=L/2$ becomes deeper and deeper with $L$.

The weight of the dip region can be estimated using (\ref{eq:dip}), as 
\beq\label{eq:TDpoidsdip}
\sum_{k=\Lambda}^{L-\Lambda} \pi(k|L)
\approx L^{-\th}
w\,c\int_{\eps}^{1-\eps}\frac{\dd\la}{[\la(1-\la)]^{1+\th}},
\eeq
where, for the sake of simplicity, we chose $\Lambda=\eps L$.
The two downhill and uphill regions are therefore well separated by the dip region, as is conspicuous on figure \ref{fig:pp1}.
Note the similarity between (\ref{eq:TDpoidsdip}) and (\ref{zrpDip}), with the correspondence
\beq\label{eq:correspond}
1-\frac{1}{n}\hookrightarrow 1-\bigmean{\frac{1}{N_L}}=w.
\eeq
 
\item
\smallskip\noindent \textit{Uphill region.}
The uphill region corresponds to $\l=L-k$ finite, where
(\ref{eq:margasym}) simplifies into
\be
\pi(L-\l|L)\approx (1-w)^2Z^{\td}(w,\l).
\ee
The weight of this region can be estimated as
\beqa\label{eq:region2}
\sum_{\l=0}^{\Lambda} \pi(L-\l|L)
&\approx& 
 (1-w)^2\sum_{\l=0}^{\Lambda} Z^{\td}(w,\l)
 \nonumber\\
 &\to&(1-w)^2\sum_{\l=0}^{\infty} Z^{\td}(w,\l)
= 1-w
\eeqa
where the last step is obtained by setting $z=1$ in the expression of the generating function (\ref{eq:fgZ}).
The right side of (\ref{eq:region2}), $1-w$, is precisely the limiting value of $\mean{1/N_L}$, see (\ref{eq:unsurN}).
This result is therefore the analogue of 
(\ref{eq:region2ZRP})
with the correspondence given in (\ref{eq:correspond}).

\end{enumerate}

In view of (\ref{eq:region1}) and (\ref{eq:region2}) we conclude that the weights of the downhill and uphill regions add to one, in line with the fact that the contribution of the dip region is subdominant, as shown above.

\subsection{Regimes for the distribution of the longest interval}
\label{sub:regimesMax}

\begin{figure}[!ht]
\begin{center}
\includegraphics[angle=0,width=.8\linewidth]{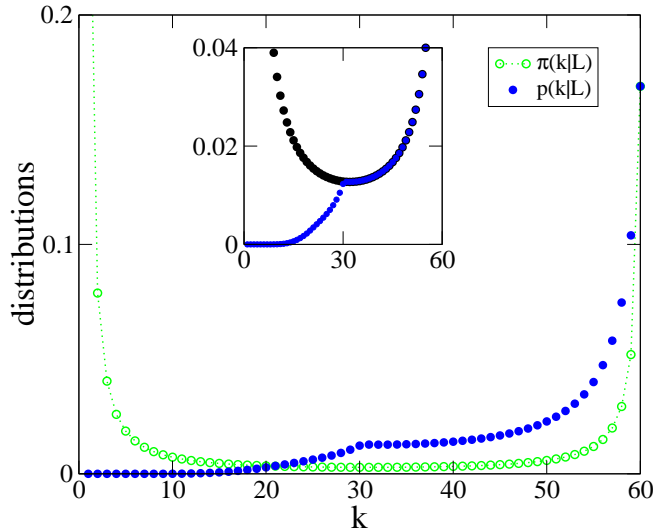}
\caption{\textsc{tdrp}: the main graph depicts the exact distributions of the size of a generic interval, $\pi(k|L)$ (in green), and of the longest one, $p(k|L)$ (in blue), in the condensed phase, for $L=60$ and $w=0.6$ computed from (\ref{eq:pkL}) and (\ref{eq:p1gf}), with \textit{Example 1} (see (\ref{eq:ex1td})).
The two curves join at $wf(L)/Z^{\td}(w,L)\approx (1-w)^2$ on the $y$-axis, for $k=60$.
The inset depicts $\pi(k|L)(1+\mean{N_{L-k}})$ (in black) and $p(k|L)$ (in blue).
For this value of $w$, the asymptotic average number $\mean{N_L}\to4$.
}
\label{fig:pp1}
\end{center}
\end{figure}

As can be seen on figure \ref{fig:pp1}, there are two main regions for the distribution of the maximum, $p(k|L)$.
For $k\le L/2$, the contribution of $p(k|L)$ to the total weight is vanishingly small.
The argument is the same as in \S\ref{sec:details}.
Hence we restrict the rest of the discussion to the region $(L/2< k\le L)$,
where $p(k|L)$ has the simpler expression $ q(k|L)$ given by (\ref{eq:qkL}).
Using (\ref{eq:ZLwlt1}), its asymptotic estimate is
\be
q(k|L)\stackunder{L\to\infty}{\approx} (1-w)^2f(k)\frac{(Z^{\td}\star Z^{\td})(w,L-k)}{f(L)},
\ee
or, equivalently, for $\l=L-k\in(0,L/2-1)$,
\beqa\label{eq:qlt1}
 q(L-\l|L)\stackunder{L\to\infty}{\approx} (1-w)^2f(L-\l)\frac{(Z^{\td}\star Z^{\td})(w,\l)}{f(L)}.
\eeqa
Let us note that the ratio
\be
r(\l)=\frac{ q(L-\l|L)}{\pi(L-\l|L)}=\frac{(Z^{\td}\star Z^{\td})(w,\l)}{Z^{\td}(w,\l)}=1+\mean{N_\l}
\ee
is an increasing function of $\l$, with first values
\be
r(0)=1,\quad r(1)=2,\quad r(2)=\frac{3w f(1)^2+2f(2)}{wf(1)^2+f(2)},
\ee
and so on,
reaching the limit, for large $\l$, 
\be
r(\l)\to \frac{2}{1-w}.
\ee
The inset in figure \ref{fig:pp1} depicts the prediction $q(k|L)=\pi(k|L)(1+\mean{N_{L-k}})$ (see (\ref{eq:better})) which coincides perfectly with $p(k|L)$ in the second half $k>L/2$.
This graph is very similar to the inset in figure \ref{fig:zrp12}.

We now discuss the behaviour of $ q(L-\l|L)$ in the two regions of interest.
\begin{enumerate}
\item For $\l$ finite, we have, see (\ref{eq:qlt1}),
\beq\label{eq:qclose}
 q(L-\l|L)\approx (1-w)^2(Z^{\td}\star Z^{\td})(w,\l).
\eeq
In particular for $\l=0$,
\be
p(L|L)=q(L|L)=\pi(L|L)=\prob(N_L=1)\approx(1-w)^2.
\ee
If $1\ll \l\ll L$, (\ref{eq:qclose}) simplifies to
\be
 q(L-\l|L)\approx \frac{2w}{1-w}f(\l).
\ee
\item
If $\l$ et $L-\l$ are simultaneously large, with $\l=\la L$, we have
\beq\label{eq:dipmax}
 q(L-\l|L)\approx\frac{2w}{1-w}f(\l)\frac{f(L-\l)}{f(L)}\approx\frac{2w}{1-w}\frac{1}{[\la(1-\la)]^{1+\th}}\frac{c}{L^{1+\th}},
\eeq
which is proportional to (\ref{eq:dip}), with ratio $2/(1-w)$.
Note the similarity of (\ref{eq:dipmax}) with (\ref{eq:zrpMaxDip}) with the correspondence
\be
n-1\hookrightarrow \mean{N_L}-1=\frac{2w}{1-w}.
\ee
The weight under the peak of the condensate tends to unity, 
\bea
\sum_{\l=0}^{\Lambda}q(L-\l|L)
&\approx& (1-w)^2\sum_{\l=0}^{\Lambda} (Z^{\td}\star Z^{\td})(w,\l)
 \nonumber\\
 &\to&(1-w)^2\sum_{\l=0}^{\infty} (Z^{\td}\star Z^{\td})(w,\l)=1,
\eea
by a computation similar to (\ref{eq:region2}), using now the fact that the last sum is equal to $1/(1-w)^2$, as can be seen by setting $z=1$ in the expression of the generating function $\tilde Z^{\td}(w,z)^2$.

\end{enumerate}

We can further analyse the fluctuations of the condensate by considering the width of the peak,
\be
L-\mean{X_\max}=\sum_{\l=0}^{L-1}\l\,q(L-\l|L)\approx \sum_{\l=0}^{L/2}\l\,q(L-\l|L).
\ee
The dominant contribution to this sum depends on whether $\th$ is smaller or larger than one.

\medskip\noindent 
$\bullet$ If $\th<1$, the dominant contribution comes from (\ref{eq:dipmax}), i.e., for $\l$ comparable to $L$,
\beqa\label{eq:correctionTdrpLt1}
L-\mean{X_\max}
&\approx& \frac{2wc}{1-w} L^{1-\th}\int_{0}^{1/2}\dd \la \frac{\la}{[(\la(1-\la)]^{1+\th}}
\nonumber\\
&\approx& \frac{2wc\,}{1-w}\mathrm{B}_{\frac{1}{2}}\Big(1-\th,-\th\Big)L^{1-\th},
\eeqa
where $\mathrm{B}(\cdot)$ is the incomplete beta function, which, for example, is equal to $2$ for $\th=1/2$.

\medskip\noindent 
$\bullet$ If $\th>1$, the main contribution comes from (\ref{eq:qclose}),
\beqa\label{eq:correctionTdrpGt1}
L-\mean{X_\max}&\approx& (1-w)^2 \sum_{\l=0}^{\Lambda}\l (Z^{\td}\star Z^{\td})(w,\l)
\nonumber\\
&\to& (1-w)^2 \sum_{\l=0}^{\infty}\l (Z^{\td}\star Z^{\td})(w,\l)
\to \frac{2w}{1-w}\xm,
\eeqa
using (\ref{eq:Z2}).
This last result (\ref{eq:correctionTdrpGt1}) has a simple interpretation.
It says that the correction $L-\mean{X_\max}$ is made of $\mean{N_L}-1=2w/(1-w)$ intervals, of mean length $\xm$.
It is therefore the perfect parallel of the result (\ref{eq:correctionzrpgt1}).
Likewise, (\ref{eq:correctionTdrpLt1}) 
is the perfect parallel of (\ref{eq:correctionzrplt1}).
In the present case the expression in the right side of this equation
is proportional to 
$\mean{N_L}-1$ times the critical mean interval $\xmtd$, given in (\ref{eq:xmtd-crit}).

\subsection{Discussion}

In view of the above analysis the following picture emerges.
The `contrast' between the dip and the condensate region increases with $L$, i.e., the dip centred around $L/2$ becomes deeper and deeper as $L^{-1-\th}$ (see (\ref{eq:dip})) relatively to the height of the peak, which is of order one.
An estimate of the contribution of the condensate to the total weight can thus be operationally obtained by summing $\pi(L-\l|L)$ for $\l$ in the layer $(0,\Lambda)$.
This sum is asymptotically equal to $1-w$, according to (\ref{eq:region2}),
which turns out to be also the asymptotic estimate of $\mean{1/N_L}$.
The interpretation of this result is clear.
When $X_1$ is larger than $L/2$, this interval is necessarily the longest one, i.e., $X_1=X_\max$.
Furthermore, since this interval is chosen amongst $N_L$ intervals, we expect that, in average,
\be
\sum_{\l=0}^{\Lambda} \pi(L-\l|L)
\approx\stackunder{1-w}{\underbrace{\bigmean{\frac{1}{N_L}}}}\
\stackunder{\approx 1}{\underbrace{\sum_{\l=0}^{\Lambda} p(L-\l|L)}}.
\ee
But the sum on the right side, namely the weight of $X_\max$ in the same layer is asymptotically equal to one.
This simple heuristic reasoning therefore recovers (\ref{eq:region2}).

In the condensed phase ($w<1, L\to\infty$) the number of intervals is finite and fluctuates around its mean, which is a superuniversal constant independent of $\th$.
This situation is akin to the case of random allocation models and \textsc{zrp}, when $L\to\infty$ and $n$ is kept fixed.
Note that, for the latter, results were independent of the value of $\th$, too.
In both situations condensation is total,
the condensed fraction is asymptotically equal to unity.

Table \ref {tab:tdrp} summarises the results found in sections \ref{sec:tdrp-disordered}, \ref{sec:tdrp-critical} and \ref{sec:tdrpcondensed}, which demonstrate a large degree of universality.

\begin{table}[htbp]
\caption{Dominant asymptotic behaviours at large $L$ for tied-down renewal processes with power-law distribution (\ref{eq:powerlaw}) for $f(k)$, in the different phases.}
\label{tab:tdrp}
\begin{center}
\begin{tabular}{|c|c||c|c||c|}
\hline
&disordered&critical $\th<1$&critical $\th>1$&condensed\\
\hline
$\mean{N_L}$&$ \frac{L}{\xmtd}$&$ L^\th$&$ \frac{L}{\xm}$&$\frac{1+w}{1-w}$\\
$\xmtd$&$\mathrm{constant}$&$ L^{1-\th}$&$\xm$&$(1-w)L$\\
$\mean{X_\max}$&$ \ln L$&$ L $&$L^{1/\th}$&$ L$\\
$Z^{\td}(w,L)$&$\e^{L/\xi} $&$ L^{\th-1} $&$\frac{1}{\xm}$&$ L^{-1-\th}$\\
\hline
\end{tabular}
\end{center}
\end{table}

\section{General statements on free renewal processes}
\label{sec:rpGeneral}

We now turn to the case of free renewal processes.
The random number $N_L$ of intervals up to $L$ is defined through the condition (\ref{eq:cond1}), $S_{N_L}<L<S_{N_L+1}$.
The size of the current (unfinished) interval---named the \textit{backward recurrence time} in renewal theory---is denoted by $B_L=L-S_{N_L}$,
see figure \ref{fig:figrenew}.
As will appear shortly, free renewal processes are more complicated to analyse than \textsc{tdrp}, essentially because now there are two kinds of intervals to consider, the intervals $X_i$ on one hand, and the last unfinished interval $B_L$, on the other hand.

In the present section, $f(k)=\prob(X=k)$ is any arbitrary distribution of the positive random variable $X$.
Later on, in sections \ref{sec:free-critical}
 and \ref{sec:rpcondensed}, $f(k)$ will obey the form (\ref{eq:powerlaw}).

\subsection{Joint distribution}

As for \textsc{tdrp} a weight $w$ is attached to each interval.
The joint probability of the configuration $\{X_1= k_1,\dots,X_{N_L}= k_n,B_L=b,N_L=n\}$, reads
\beqa
p(k_1,\ldots,k_n,b,n|L)
&=&\prob(\{X_i=k_i\},B_L=b,N_L=n)
\nonumber\\
&=&\frac{1}{Z^{\free}(w,L)} w^nf(k_1)\ldots f(k_n)\, g(b)\,\delta\big(\sum_{i=1}^n k_i+b,L\big),
\label{eq:jointerp}
\eeqa
where $ g(b)$ is the tail probability (or complementary distribution function) defined in (\ref{eq:defg}),
\beq\label{eq:qb}
 g(b)=\prob(X>b)=\sum_{k> b} f( k)=f(b+1)+f(b+2)+\cdots
\eeq
As mentioned in section \ref{sec:tdrpGeneral}, since the summands $X_i$ have the interpretation of the sizes of the intervals,
 we take $f(0)=0$.
For $n=0$,
\be
p(\{\},b,0|L)\num= g(b)\delta(b,L),
\ee
corresponding to the event of no renewal occurring between $0$ and $L$, i.e., $B_L=L$, and
where $\{\}$ means empty.
The generating function of $g(b)$ is
\beq\label{eq:qtilde}
\tilde g(z)
=\sum_{b\ge0}z^b g(b)
=\frac{1}{1-z}-\sum_{b\ge1}z^b\sum_{ k=1}^bf( k)
=\frac{1-\tilde f(z)}{1-z},
\eeq
with $\tilde g(0)= g(0)=1$.

The denominator of (\ref{eq:jointerp}) is the free partition function, obtained by summing on $n$, on the $k_i$ and on $b$,
\beqa 
Z^{\free}(w,L)&=&\sum_{n\ge0}\sum_{b\ge0}\sum_{\{k_i\}}\,p(k_1,\ldots, k_n,b,n|L)
\nonumber\\ 
&=&\sum_{n\ge0}w^n\sum_{b\ge0}\sum_{\{k_i\}}f(k_1)\cdots f(k_n) g(b)\delta\big(\sum_{i=1}^n k_i+b,L\big)
\nonumber\\ 
&=&\sum_{b\ge0} g(b)\Big[\delta(b,L)+\sum_{k_1}wf(k_1)\delta(k_1+b,L)
\nonumber\\ 
&+&\sum_{k_1,k_2}w^2f(k_1)f(k_2)\delta(k_1+k_2+b,L)
+\dots\Big]
\nonumber\\ 
&=&\,
\stackunder{n=0}{\underbrace{ g(L)}}+\stackunder{n=1}{\underbrace{w\,g\star f}}+
\stackunder{n=2}{\underbrace{w^2\,g\star f\star f}}+\dots
=\sum_{n\ge0}w^n\big(g\star(f\star)^n\big)(L).
\label{eq:rpZwL}
\eeqa
For instance, 
\bea
Z^{\free}(w,0)=1,\qquad
Z^{\free}(w,1)= g(1)+wf(1),\qquad
\\
Z^{\free}(w,2)=g(2)+wf(2)+wf(1) g(1)+w^2f(1)^2,
\eea
and so on.
From (\ref{eq:rpZwL}) we have
\beq\label{eq:Z}
\tilde Z^{\free}(w,z)=\sum_{L\ge0}z^L Z^{\free}(w,L)=\sum_{n\ge0}\big(w\tilde f(z)\big)^n\tilde g(z)=
\frac{\tilde g(z)}{1-w\tilde f(z)}.
\eeq
Let us note that
\beq\label{eq:convol}
Z^{\free}(w,L)=\sum_{n\ge0}w^n\sum_{b\ge0}\prob(S_n=L-b) g(b)
=(g\star Z^{\td})(w,L),
\eeq
where $Z^{\td}(w,L)$ is the partition function (\ref{eq:ZwL}) for \textsc{tdrp}.

The case of usual (unweighted) free renewal processes is recovered by setting $w=1$ in these expressions.
This yields $Z^{\free}(1,L)=1$, as can be seen from (\ref{eq:Z}), and the joint probability distribution (\ref{eq:jointerp}) simplifies accordingly.

\subsection{Distribution of the number of intervals}
As for \textsc{tdrp} we denote this distribution as%
\footnote{Whenever no ambiguity arises, we use the same notations for the observables of the tied-down and free renewal processes.
Otherwise, when necessary, we add a superscript, as e.g., for $Z^{\td}$, $Z^{\free}$ or in (\ref{eq: NLfree}).}
\be
\pp_n(L)=\prob(N_L=n).
\ee
We read on the successive terms of $Z^{\free}(w,L)$ that
\beq\label{eq:probN0}
\pp_0(L)=\frac{ g(L)}{Z^{\free}(w,L)}=\prob(B_L=L),\quad
\pp_1(L)=\frac{w(g\star f)(L)}{Z^{\free}(w,L)},
\eeq
and so on.
More generally,
\be
\pp_n(L)=\frac{w^n\big(g\star(f\star)^n\big)(L)}{Z^{\free}(w,L)}=\frac{w^n(g\star Z_n)(L)}{(g\star Z^{\td})(w,L)}.
\ee
Summing (\ref{eq:jointerp}) on $b$ and on the $k_i$, and taking the generating function with respect to $L$ yields
\be
\sum_{L\ge0}z^L\pp_n(L)\num=\Big(w\tilde f(z)\Big)^n\tilde g(z),
\ee
to be compared to (\ref{eq:Z}).
Therefore
\beq\label{eq:fgNfree}
\sum_{L\ge0}z^L\mean{N_L}\num=\frac{w\tilde f(z)\tilde g(z)}{(1-w\tilde f(z))^2}=w\frac{\dd}{\dd w}\tilde Z^{\free}(w,z),
\eeq
and
\beq\label{eq:NavRP}
\mean{N_L}=w\frac{\dd\ln Z^{\free}(w,L)}{\dd w},
\eeq
as for \textsc{tdrp}, see (\ref{eq:meanNL}).
More generally,
\be
\sum_{L\ge0}z^L\bigmean{v^{N_L}}\num=\sum_{n\ge0}v^n\Big(w\tilde f(z)\Big)^n\tilde g(z)=\tilde Z^{\free}(vw,z),
\ee
so we obtain, as for \textsc{tdrp}, see (\ref{eq:yNL}),
\beq\label{eq:vNLrenew}
\bigmean{v^{N_L}}=\sum_{n\ge0}v^n \pp_n(L)=\frac{Z^{\free}(vw,L)}{Z^{\free}(w,L)}.
\eeq
Finally, comparing (\ref{eq:fgNfree}) to (\ref{eq:NLfg}), we note the relationship between the free and tied-down cases,
\beq\label{eq: NLfree}
\mean{N^{\free}_L}\num=\sum_{k=0}^Lg(k)\mean{N^{\td}_{L-k}}\num,
\eeq
hence using (\ref{eq:niqTdc}),
we have
\be
\mean{N^{\free}_L}\num=(g\star Z^{\td}\star Z^{\td})(w,L)-Z^{\free}(w,L),
\ee
or, equivalently,
\beq\label{eq:RPidentity}
(g\star Z^{\td}\star Z^{\td})(w,L)=Z^{\free}(w,L)(1+\mean{N_L}).
\eeq

\subsection{Distribution of $S_{N_L}$}
We recall that this quantity is the sum
of the $N_L$ intervals before $L$, see (\ref{eq:SNL}) and figure \ref{fig:figrenew}.
By definition,
\beq
\prob(S_{N_L}=j)=\mean{\delta(S_{N_L},j)}
=\sum_{n\ge0}\sum_{b\ge0}\sum_{\{k_i\}}\delta(S_{N_L},j)p(k_1,\ldots,k_n,b,n|L).
\eeq
Thus, using (\ref{eq:jointerp}),
we have
\be
\sum_{L\ge0}z^L\sum_{j=0}^L x^{j} \prob(S_{N_L}=j)\num=\frac{\tilde g(z)}{1-w\tilde f(xz)},
\ee
which generalises the expression for this quantity when $w=1$ \cite{gl2001}.
By derivation with respect to $x$ then setting $x=1$, leads to
\beq\label{eq:Smoy}
\sum_{L\ge0}z^L\mean{S_{N_L}}\num=\frac{wz\tilde f'(z)\tilde g(z)}{(1-w\tilde f(z))^2},
\eeq
whose summation with (\ref{eq:numBfg}) below leads to the equality
\be
\sum_{L\ge0}z^L\big(\mean{S_{N_L}}\num+\mean{B_L}\num\big)
=z\frac{\dd \tilde Z^{\free}(w,z)}{\dd z},
\ee
which expresses the sum rule
\be
\mean{S_{N_L}}+\mean{B_L}=L.
\ee
The asymptotic behaviours of these quantities for $w=1$ are simple \cite{gl2001}.
If $\th<1$, then $\mean{S_{N_L}}\approx \th L$, $\mean{B_L}\approx (1-\th) L$.
If $1<\th<2$, then $\mean{S_{N_L}}\approx L-cL^{2-\th}/[\th(2-\th)\xm]$, and $\mean{B_L}$ follows by difference.

\subsection{Distribution of $B_L$}

The distribution of $B_L$ is obtained by summing (\ref{eq:jointerp}) on the $k_i$ and on $n$
\be 
\prob(B_L=b)=\frac{1}{Z^{\free}(w,L)} g(b)\sum_{n\ge0}w^n\sum_{\{ k_i\ge1\}}f( k_1)\cdots f( k_n)\delta\Big(\sum_{i=1}^n k_i+b,L\Big).
\ee
This entails that 
\beq\label{eq:probB} 
\prob(B_L=b)\num=
g(b)\sum_{n\ge0} w^n\prob(S_n=L-b) 
=g(b)Z^{\td}(w,L-b) ,
\eeq
hence, using (\ref{eq:convol})
\be
\prob(B_L=b)=\frac{g(b)Z^{\td}(w,L-b)}{(g\star Z^{\td})(w,L)}.
\ee
The generating function with respect to $L$ of (\ref{eq:probB}) reads
\beq\label{eq:numB}
\sum_{L\ge0}z^L\prob (B_L=b)\num
=\frac{z^b g(b)}{1-w\tilde f(z)},
\eeq
which summed upon $b$ gives $\tilde Z^{\free}(w,z)$ back.
Taking now the generating function of (\ref{eq:numB}) with respect to $b$ yields
\be
\sum_{L\ge0}z^L\sum_{b\ge0}y^b\prob (B_L=b)\num
=\frac{\tilde g(yz)}{1-w\tilde f(z)}.
\ee
The mean $\mean{B_L}$ ensues by taking the derivative of the right side of this equation with respect to $y$ and setting $y$ to one,
\beq\label{eq:numBfg}
\sum_{L\ge0}z^L\mean{B_L}\num=\frac{z\tilde g'(z)}{1-w\tilde f(z)}.
\eeq

\subsection{Single interval distribution}

As for \textsc{tdrp}, the single interval distribution,
\be
\pi(k|L)=\mean{\delta(X_1, k)},
\ee
is obtained by summing $p( k_1,\dots, k_n,b, n|L)$ on $k_1,\dots, k_n$, $b$ and $n\ge1$
\beqa \label{eq:marginalpkL}
\pi(k|L)\num
&=&\sum_{b\ge0} g(b)\Big[
\sum_{ k_1}\delta( k_1, k)w f( k_1)\delta( k_1+b,L)
\nonumber\\
&+&\sum_{ k_1, k_2}\delta( k_1,k)w^2f( k_1)f( k_2)\delta( k_1+ k_2+b,L)
+\cdots\Big]
\nonumber\\
&=&\sum_{b\ge0} g(b)\Big[wf(k)\delta(k+b,L)
+\sum_{k_2}w^2f(k)f(k_2)\delta(k+k_2+b,L)+\dots\Big]
\nonumber\\
&=&wf(k)Z^{\free}(w,L-k).
\eeqa
So
\beq\label{eq:rppkLres}
\pi(k|L)=\frac{wf(k)Z^{\free}(w,L-k)}{Z^{\free}(w,L)}.
\eeq
The generating function of the numerator is therefore
\beqa\label{eq:fgmarg}
\sum_{L\ge0}z^L\pi(k|L)\num
=wz^kf(k)\tilde Z^{\free}(w,z)=\frac{wz^kf(k)\tilde g(z)}{1-w\tilde f(z)},
\eeqa
to be compared to (\ref{eq:fgtaum}).
Though (\ref{eq:rppkLres}) is formally identical to (\ref{eq:pkL})
the normalisations of these two distributions are different.
Indeed,
\bea 
\sum_{L\ge0}z^L\sum_{k=1}^L\pi(k|L)\num
=w\tilde f(z)\tilde Z^{\free}(w,z)
=\frac{w\tilde f(z)\tilde g(z)}{1-w\tilde f(z)}=\tilde Z^{\free}(w,z)-\tilde g(z),
\eea
which means that
\beq\label{eq:sumrule}
1-\sum_{k=1}^L\pi(k|L)
=\frac{ g(L)}{Z^{\free}(w,L)}=\pp_0(L)=\prob(B_L=L).
\eeq
In other words
\be
\sum_{k=1}^L\prob(X=k|L)+\prob(B_L=L)=1.
\ee
The distribution $\pi(k|L)$ is thus defective.
The recursion relation for $Z^{\free}(w,L)$ follows from (\ref{eq:marginalpkL}) and (\ref{eq:sumrule})
\beq\label{eq:rprecursionZ}
Z^{\free}(w,L)= g(L)+\sum_{k=1}^Lwf(k)Z^{\free}(w,L-k).
\eeq
At the end point, $k=L$, we have
\beq\label{eq:terminal+}
\pi(L|L)=\frac{wf(L)}{Z^{\free}(w,L)},
\eeq
which corresponds to the event $\{X_1=L,B_L=0,N_L=1\}$, and which is formally the same as (\ref{eq:pLLtdrp}).

Both sides of (\ref{eq:rprecursionZ}) are equal to unity if $w=1$.
Note that the computation of $\pi(k|L)$ for $w=1$ is given in \cite{gl2001}, yielding $\pi(k|L)=f(k)$ with $k\le L$, which shows that this distribution is already defective for $w=1$, with $\sum_{k=1}^L\pi(k|L)=1-g(L)$.

\subsection{Mean interval $\xmtd$}
We proceed as in \S \ref{sub:meaninterval}.
The mean interval is, by definition,
\be
\xmtd=\sum_{k\ge1}k\pi(k|L).
\ee
Multiplying (\ref{eq:fgmarg}) by $k$ and summing upon $k$ yields
\beq\label{eq:fgXmoyen}
\sum_{L\ge1}z^L
\xmtd \num=\frac{w z\tilde f'(z)\tilde g(z)}{1-w \tilde f(z)},
\eeq
to be compared to (\ref{eq:fgtaum}) for \textsc{tdrp}.

\subsection{The longest interval}
In the present case the longest interval is defined as
\be
X_\max=\max(X_1,X_2,\dots,B_L).
\ee
Its distribution function is 
\be 
F(k|L)=\prob(X_{\max}\le k|L)
=\sum_{n\ge0}\sum_{b=0}^k\sum_{k_1=1}^{k}\dots\sum_{k_n=1}^{k}p(\{k_i\},b,n|L)
=\frac{F(k|L)\num}{Z^{\free}(w,L)},
\ee
with initial value
\be
F(k|0)\num=1.
\ee
As for \textsc{tdrp}, $F(L|L)\num=Z^{\free}(w,L)$, hence $F(L|L)=1$.
The generating function of the numerator is
\beqa 
\sum_{L\ge0}z^L F(k|L)\num
&=&\tilde g(z,k)\Bigg(1+\sum_{n\ge1}\prod_{i=1}^n\Big(\sum_{k_i=1}^{k} wf(k_i)z^{k_i}\Big)\Bigg)
\nonumber\\ 
&=&\tilde g(z,k)\Bigg(1+\sum_{n\ge1}\Big(w\tilde f(z,k)\Big)^n\Bigg)
\label{eq:fgF1Nrp}
=\frac{\tilde g(z,k)}{1-w\tilde f(z,k)},
\eeqa
where
\be
\tilde f(z,k)=\sum_{j=1}^{k} z^{j} f(j),\qquad \tilde g(z,k)=\sum_{b=0}^k z^b g(b),
\ee
are related by 
\be
1-\tilde f(z,k)=z^{k+1}g(k)+(1-z)\tilde g(z,k).
\ee
Note that
\beqa\label{eq:FfFtd}
F^{\free}(k|L)\num&=&\sum_{b=0}^kg(b)F^{\td}(k|L-b)\num
\nonumber\\
&=&\sum_{b=0}^kg(b)\sum_{n=0}^Lw^nF_n(k|L-b)\num,
\eeqa
where we used (\ref{eq:F1klnum}) in the last step.

The distribution of $X_\max$ is given by the difference
\be
p(k|L)=\prob(X_{\max}=k)=F(k|L)-F(k -1|L),
\ee
where $F(0|L)=\delta(L,0)$,
with generating function
\bea
\sum_{L\ge0}z^Lp(k|L)\num
&=&\tilde g(z,k)\Big(\frac{1}{1-w\tilde f(z,k)}-\frac{1}{1-w\tilde f(z,k-1)}\Big)
\nonumber\\
&=&\frac{wz^{k}f(k)\tilde g(z,k)}{[1-w\tilde f(z,k)][1-w\tilde f(z,k-1)]}.
\eea
At the end point, $k=L$, we have
\beq\label{eq:rpp1LL}
p(L|L)=\frac{wf(L)+g(L)}{Z^{\free}(w,L)}=\pi(L|L)+\prob(B_L=L),
\eeq
where the last two terms correspond respectively to the events $\{X_1=L,B_L=0,N_L=1\}$, cf (\ref{eq:terminal+}) and 
$\{B_L=L,N_L=0\}$, cf (\ref{eq:probN0}).
The mean is given by the sum
\be
\mean{X_\max}=\sum_{k=0}^L\Big(1-F(k|L)\Big)=L-\sum_{k=1}^{L-1}F(k|L),
\ee
which implies a relation between the generating functions
\bea
\sum_{L\ge0}z^L{\mean{X_\max}}\num=\sum_{k\ge0}\Big(\tilde Z^{\free}(w,z)-\tilde F(k,z)\num\Big)
\\
=\sum_{k\ge0}\Big(\frac{\tilde g(z)}{1-w\tilde f(z)}-\frac{\tilde g(z,k)}{1-w\tilde f(z,k)}\Big),
\eea
where $\tilde F(k,z)\num$ is given by (\ref{eq:fgF1Nrp}).

Denoting again the restriction of $p(k|L)$ to $k>L/2$ by $q(k|L)$, we can obtain an expression of this quantity by a similar reasoning as done for (\ref{eq:qkL}) in \S \ref{sub:longestTDRP}.
We thus obtain
\beqa\label{eq:qkLRenew} 
q(k|L)&=&\prob(B_L=k)+\frac{wf(k)}{Z^{\free}(w,L)}\sum_{b=0}^{L-k}g(b)(Z^{\td}\star Z^{\td})(w,L-k-b)
\nonumber\\ 
&=&\prob(B_L=k)+\frac{wf(k)\big(g\star Z^{\td}\star Z^{\td}\big)(w,L-k)}{Z^{\free}(w,L)},
\eeqa
expressing the fact that the longest interval can be either $B_L$ or a generic interval $X_i$.
Equivalently, using (\ref{eq:RPidentity}), this reads
\beqa\label{eq:beautiful}
q(k|L)=\prob(B_L=k)+\frac{wf(k)Z^{\free}(w,L-k)}{Z^{\free}(w,L)}(1+\mean{N_{L-k}}),
\eeqa
or else
\beqa\label{eq:beautiful+}
q(k|L)=\prob(B_L=k)+\pi(k|L)(1+\mean{N_{L-k}}),
\eeqa
to be compared to (\ref{eq:tdc}) and (\ref{eq:better}). 
For $k=L$, (\ref{eq:rpp1LL}) is recovered.

 \subsection*{Remarks} 

1. We can prove (\ref{eq:qkLRenew}) otherwise.
We start from the first line of (\ref{eq:FfFtd}) and take the discrete derivative with respect to $k$ of both sides, which yields
\be
 \Delta_k F^{\free}(k|L)\num=g(k)F^{\td}(k|L-k)\num+\sum_{b=0}^k g(b)\Delta_kF^{\td}(k|L-b)\num.
\ee
If $k>L/2$, using (\ref{eq:FtdZtd}), we recognise
the numerator of the first term of (\ref{eq:qkLRenew}) in the first term of this equation.
Likewise, if $k>L/2$, it can be shown that the second term of the above equation is equal to the numerator of the second term of (\ref{eq:qkLRenew}).

\smallskip
\noindent 2.
From (\ref{eq:FtdZtd}) we infer that, if $k>L/2$,
\be
 F^{\free}(k|L-k)\num=Z^{\free}(w,L-k).
\ee

Lastly, consider the probability ${Q_L}$ that the last unfinished interval is the longest one, 
that is
\beq\label{eq:defQ}
{Q_L}=\prob (B_L\ge\max(X_1,\dots,X_{N_L}))
=\sum_{n\ge0}\sum_{b\ge0}\sum_{k_1=1}^{b}\dots\sum_{k_n=1}^{b}p(\{k_i\},b,n|L).
\eeq
The generating function with respect to $L$ of its numerator reads
\be
\sum_{L\ge0}z^L{Q_L}\num=\sum_{b\ge0}\frac{z^bg(b)}{1-w\tilde f(z,b)},
\ee
hence
\bea
\tilde Z^{\free}(w,z)-\sum_{L\ge0}z^L{Q_L}\num
&=&\frac{\tilde g(z)}{1-w\tilde f}-\sum_{b\ge0}\frac{z^bg(b)}{1-w\tilde f(z,b)}
\\
&=&\sum_{b\ge0}z^bg(b)\Big(\frac{1}{1-w\tilde f}-\frac{1}{1-w\tilde f(z,b)}\Big).
\eea

\section{Critical regime ($w=1$) for free renewal processes}
\label{sec:free-critical}
In this section and in the following one (section \ref{sec:rpcondensed})
we specialise the discussion to the case of a subexponential
distribution $f(k)=\prob(X=k)$ with asymptotic power-law decay (\ref{eq:powerlaw}).

The critical regime is thoroughly described in \cite{gl2001,gms2015} and builds upon previous studies \cite{feller,dynkin,lamperti58,lamperti61}.
The results are summarised in table \ref {tab:renewal}, which also presents the main outcomes for the disordered regime ($w>1$).

The initial analysis of the distribution of the longest interval is due to Lamperti \cite{lamperti61}.
Let us just recover the universal asymptotic expression of $q(k|L)$, for $1\ll k \sim L$, with $r=k/L$ fixed, when $\th<1$, given in \cite{lamperti61},
\beq\label{eq:lamperti}
q(k|L)\approx \frac{1}{L}\frac{\sin\pi\th}{\pi}\frac{1}{r^{1+\th}(1-r)^{1-\th}}.
\eeq
This result can be simply inferred from (\ref{eq:beautiful+}).
For the first term of (\ref{eq:beautiful+}), we obtain for $L$ large, using (\ref{eq:Z1asympt}),
\be
\prob(B_L=k)=g(k)Z^{\td}(1,L-k)\stackunder{L\to\infty}{\approx} \frac{\sin\pi \th}{\pi}\frac{1}{k^\th(L-k)^{1-\th}},
\ee
i.e., an arcsine law in the variable $r=k/L$, which is a well-known result \cite{dynkin,feller,gl2001}.
For the second term of (\ref{eq:beautiful+}), we need 
\be
\mean{N_{L-k}}\stackunder{L\to\infty}{\approx}\frac{\sin\pi \th}{\pi c}(L-k)^\th,
\ee
which is obtained using (\ref{eq:fgNfree}).
Adding the two terms of (\ref{eq:beautiful+}) yields (\ref{eq:lamperti}).

\section{Condensed phase ($w<1$) for free renewal processes}
\label{sec:rpcondensed}

We now focus on the case of most interest, namely the condensed phase, $w<1$, for subexponential distributions (\ref{eq:powerlaw}).
As in section \ref{sec:tdrpcondensed}, we investigate the statistics of the number of intervals, the single-interval distribution and we characterise the fluctuations of the condensate.
We also address the statistics of the last interval $B_L$.

\subsection{Asymptotic estimates at large $L$}

The asymptotic analysis of (\ref{eq:Z}) yields, for large $L$,
\beqa\label{eq:rpZasymplt1}
Z^{\free}(w,L)&\approx& \frac{g(L)}{1-w}\Big(1-\frac{wc\Gamma(1-\th)^2}{(1-w) \Gamma(1-2\th)\th L^\th}\Big),
\quad \th<1
\\
Z^{\free}(w,L)&\approx& \frac{g(L)}{1-w}\Big(1+\frac{2w\th\xm}{(1-w)L}\Big),\quad \th>1.
\label{eq:rpZasympgt1}
\eeqa
where $g(\cdot)$ is defined in (\ref{eq:defg}).
As a consequence, (\ref{eq:vNLrenew}) yields
\be
\bigmean{v^{N_L}}\stackunder{L\to\infty}{\to}\frac{1-w}{1-vw},
\ee
leading asymptotically to a geometric distribution for $N_L$, 
\beq\label{eq:geo-free}
\prob(N_L=n)\to \pp_n= (1-w)w^n,
\eeq
independent of $\th$, from which entails, in the same limit,
\beq\label{eq:renewN}
\mean{N_L}\to\frac{w}{1-w}.
\eeq
For $L$ large but finite, we find, using (\ref{eq:NavRP}) and (\ref{eq:rpZasymplt1}) or (\ref{eq:rpZasympgt1}),
\beqa\label{eq:NavasympRP}
\mean{N_L}\approx \frac{w}{1-w}\Big(1-\frac{c}{1-w}\frac{\th\,\Gamma(1-\th)^2}{\Gamma(1-2\th)L^\th}\Big),
\quad \th<1.
\nonumber\\
\mean{N_L}\approx \frac{w}{1-w}\Big(1+\frac{2\th\xm}{(1-w)L}\Big),
\quad \th>1.
\eeqa

The asymptotic estimate of the mean interval can be obtained from the analysis of (\ref{eq:fgXmoyen}),
\bea
\xmtd&\approx& \frac{wc\Gamma(1-\th)^2}{\Gamma(2-2\th)}L^{1-\th},\quad \th<1,
\\
\xmtd&\approx& w(1+\th)\xm,\quad \th>1.
\eea
These expressions can also be obtained by means of the marginal distribution $\pi(k|L)$, see below.
Likewise, the asymptotic estimate of the mean sum can be obtained from the analysis of (\ref{eq:Smoy}),
\bea
\mean{S_{N_L}}&\approx&\frac{wc}{1-w}\frac{\Gamma(1-\th)^2}{\Gamma(2-2\th)}L^{1-\th},\qquad \th<1,
\\
\mean{S_{N_L}}&\approx&\frac{w}{1-w}(1+\th)\xm,\qquad \th>1,
\eea
hence
\beqa\label{eq:correctionBlt1}
\mean{B_L}&\approx& L-\frac{wc}{1-w}\frac{\Gamma(1-\th)^2}{\Gamma(2-2\th)}L^{1-\th},\qquad \th<1,
\\
\mean{B_L}&\approx& L-\frac{w}{1-w}(1+\th)\xm,\qquad \th>1.
\label{eq:correctionBgt1}
\eeqa

\subsection{Regimes for the single interval distribution}

\begin{figure}[!ht]
\begin{center}
\includegraphics[angle=0,width=.8\linewidth]{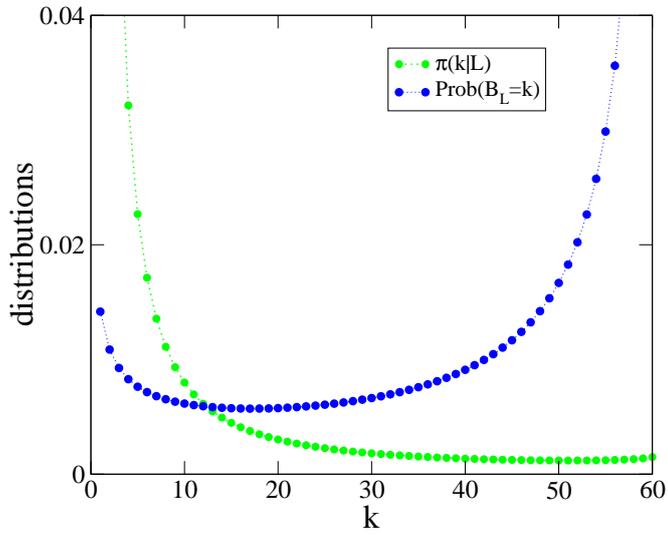}
\caption{Free renewal processes: distributions $\pi(k|L)$ (in green) and $\prob(B_L=k)$ (in blue) for \textit{Example 1} (see (\ref{eq:ex1td})), with $w=0.8$ and $L=60$. 
For this value of $w$, the asymptotic average number $\mean{N_L}\to4$.
}
\label{fig:B-marg}
\end{center}
\end{figure}

We proceed as was done for \textsc{tdrp} (see section \ref{sub:tdrpregimes}).
Using (\ref{eq:rpZasymplt1}) we have the estimate, at large $L$,
\beq\label{eq:rppkL}
\pi(k|L)\stackunder{L\to\infty}{\approx} w(1-w)\frac{f(k)Z^{\free}(w,L-k)}{g(L)}.
\eeq
Figure \ref{fig:B-marg} depicts the distribution $\pi(k|L)$ (together with the distribution $\prob(B_L=k)$, see \S \ref{sub:regimesBL} below), for $L=60$ and $w=0.8$ computed with \textit{Example 1} (see (\ref{eq:ex1td})).
As can be seen on this figure, there are three distinct regions for $\pi(k|L)$, that we consider in turn.

\begin{enumerate}
\item \textit{Downhill region.}
For $k$ finite, using (\ref{eq:rpZasymplt1}) again, we have
\be
\pi(k|L)\approx \frac{wf(k)g(L-k)}{g(L)}\approx wf(k).
\ee
\item \textit{Dip region.}
When $k$ and $L-k$ are simultaneously large, setting $k=\la L$ in (\ref{eq:rppkL}) 
($0<\la<1$) yields the estimate, at large $L$,
\beq\label{eq:diprp} 
\pi(k|L)\approx \frac{wf(k)g(L-k)}{g(L)} 
\approx \frac{1}{\la^{1+\th}(1-\la)^\th}\frac{wc}{L^{1+\th}}
\approx \frac{1}{\la^{1+\th}(1-\la)^\th}wf(L),
\eeq
with a dip centred around $k_{\min}=L(1+\th)/(1+2\th)$.
\item
In the region corresponding to $L-k$ finite, 
(\ref{eq:rppkL}) simplifies into
\be
\pi(k|L)\approx \frac{w(1-w)\th}{L}Z^{\free}(w,L-k).
\ee
In particular, for $k=L$, $\pi(L|L)\approx w(1-w)\th/L$, cf (\ref{eq:terminal+}).

\end{enumerate}

The weight of the downhill region is found to be asymptotically equal to $w$, using the same reasoning as in \S \ref{sub:tdrpregimes}.
The complement is borne by $g(L)/Z^{\free}(w,L)\approx 1-w$, see (\ref{eq:sumrule}) and (\ref{eq:sumrule+}).
The two other regions therefore do not contribute to the total weight, asymptotically.

In order to complete the picture we now investigate the distribution of $B_L$, the last unfinished interval.

\subsection{Regimes for the distribution of $B_L$}
\label{sub:regimesBL}

According to (\ref{eq:probB}), and in view of (\ref{eq:rpZasymplt1}), for $L$ large we have
\beq\label{eq:rpProbB}
\prob(B_L=b)\approx (1-w) \frac{ g(b)Z^{\td}(w,L-b)}{g(L)}.
\eeq
Let us discuss the different regimes of this expression according to the magnitude of $b$.

\begin{enumerate}
\item If $b$ is finite, the asymptotic estimate of $Z^{\td}(w,L-b)$ is given by (\ref{eq:ZLwlt1}), hence (\ref{eq:rpProbB}) becomes
\beq\label{eq:rpregimebfini}
\prob(B_L=b)\approx \frac{w}{1-w}\frac{ g(b)f(L-b)}{ g(L)}.
\eeq

\item
If $b\sim L$, the same estimate (\ref{eq:rpregimebfini}) still holds, then setting $L-b=\la L$, we get
\beq\label{eq:rpdip}
\prob(B_L=b)\approx \frac{w}{1-w}\frac{ g(b)f(L-b)}{ g(L)}
\approx \frac{wc}{1-w}\frac{L^{-1-\th}}{\la^{1+\th}(1-\la)^{\th}},
\eeq
which has its minimum at $k_{\min}=L\,\th/(1+2\th)$.
\item
If $L-b$ is finite, (\ref{eq:rpProbB}) becomes
\beq\label{eq:pBLiii}
\prob(B_L=b)\approx (1-w)Z^{\td}(w,L-b),
\eeq
in particular, see (\ref{eq:sumrule}),
\beq\label{eq:sumrule+}
\prob(B_L=L)=\pp_0(L)=\frac{ g(L)}{Z^{\free}(w,L)} \to 1-w.
\eeq

\end{enumerate}

Let us estimate, for later use, the probability that $B_L$ is less than $L/2$.
The result depends on the value of $\th$.

\noindent$\bullet$
If $\th<1$, using (\ref{eq:rpdip}), we have
\beq\label{eq:Blthalf}
\prob(B_L\le L/2)
\approx
\frac{wc}{1-w}L^{-\th}\int_{1/2}^1\frac{\dd\la}{\la^{1+\th}(1-\la)^{\th}}
\approx \frac{wc}{1-w}\mathrm{B}_{\frac{1}{2}}\Big(1-\th,-\th\Big)L^{-\th}.
\eeq

\noindent$\bullet$
If $\th>1$, using (\ref{eq:rpregimebfini}), we have 
\beq\label{eq:Blthalf+}
\prob(B_L\le L/2)
\approx
\frac{w\,\th}{(1-w)L} \sum_{b=0}^{L/2} g(b)
\approx \frac{w\,\th}{(1-w)L} \sum_{b=0}^{L} g(b)
\approx \frac{w\,\th \xm}{(1-w)L}.
\eeq

%
\begin{figure}[!ht]
\begin{center}
\includegraphics[angle=0,width=.8\linewidth]{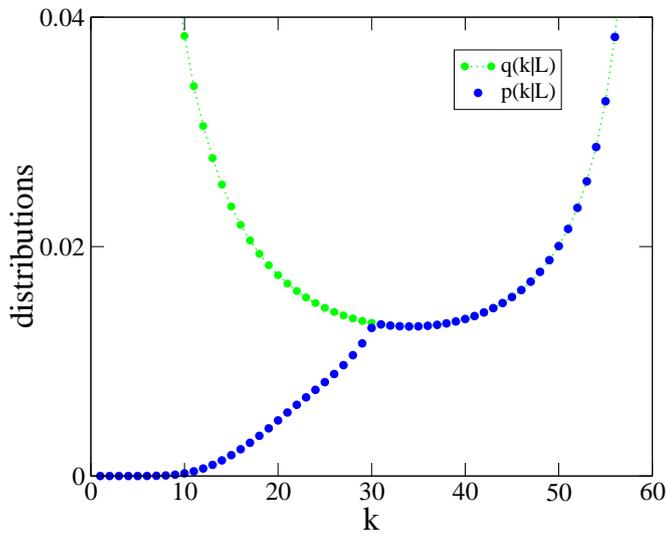}
\caption{Free renewal processes: distributions $\prob(B_L=k)+\pi(k|L)(1+\mean{N_{L-k}})$ (in green) and $p(k|L)$ (in blue) for \textit{Example 1} (see (\ref{eq:ex1td})), with $w=0.8$ and $L=60$. 
The distribution $q(k|L)$ is the restriction to $k>L/2$ of $\prob(B_L=k)+\pi(k|L)(1+\mean{N_{L-k}})$ according to (\ref{eq:beautiful+}).
For this value of $w$, the asymptotic average number $\mean{N_L}\to4$.
This figure is to be compared to the insets of figures \ref{fig:zrp12} and \ref{fig:pp1}.
}
\label{fig:pp1Free}
\end{center}
\end{figure}

%
\subsection{Regimes for the distribution of the longest interval}
\label{sub:RPlongest}

The bulk of the distribution of $X_\max$ lies in the region $k>L/2$, and is therefore given by $q(k|L)$, see (\ref{eq:qkLRenew}) or (\ref{eq:beautiful+}).

We start by giving an illustration.
The distribution $p(k|L)$ of $X_\max$ on the whole interval $(0,L)$ is depicted in figure \ref{fig:pp1Free} for \textit{Example 1} (see (\ref{eq:ex1td})).
The restriction of this distribution to the second half $k>L/2$, that is $q(k|L)$, computed from (\ref{eq:beautiful+}) is also depicted.
The corresponding figure for $\th>1$ is qualitatively alike.

Let us now compare the respective contributions of each of the two terms in (\ref{eq:qkLRenew}) to the total weight in the region $k>L/2$.
The first term is investigated in \S \ref{sub:regimesBL} above.
The asymptotic estimate of the second term is as follows.
We start with the case $\th<1$.

\medskip\noindent 
$\bullet$ If $\th<1$, we consider two regimes.\\
\smallskip\noindent {\bf (i)} The main contribution of the second term to the total weight comes from the regime $L-k\sim L$.
Using the asymptotic estimate for large $L$,
\beq\label{eq:discrep}
\big(g\star Z^{\td}\star Z^{\td}\big)(w,L)\approx
\frac{g(L)}{(1-w)^2}\Big(1-\frac{2wc}{1-w}\frac{\th\Gamma(-\th)^2}{\Gamma(1-2\th)L^\th}\Big),
\eeq
and setting $L-k=\la L$, the second term reads
\beq\label{eq:terme2as}
\frac{wf(k)}{Z^{\free}(w,L)}\frac{c}{(1-w)^2\th (L-k)^\th}
\approx 
\frac{wc}{1-w}\frac{L^{-1-\th}}{\la^{\th}(1-\la)^{1+\th}},
\eeq
which is similar to (\ref{eq:rpdip}).
Summing this expression upon $\la$ from $0$ to $1/2$ yields (\ref{eq:Blthalf}), that is $\prob(B_L\le L/2)$.
Therefore adding this contribution to the first one, namely $\prob(B_L>1/2)$,
gives unity, up to small corrections, in agreement with the fact that the weight of $p(k|L)$ in the left domain $k<L/2$ is negligible.

\smallskip\noindent {\bf (ii)} If $L-k$ is finite, then the second term reads
\be
\frac{wf(k)\big(g\star Z^{\td}\star Z^{\td}\big)(w,L-k)}{Z^{\free}(w,L)}
\approx \frac{(1-w)w\th}{L}\big(g\star Z^{\td}\star Z^{\td}\big)(w,L-k),
\ee
which is subdominant compared to (\ref{eq:pBLiii}).

In order to get the correction of $\mean{X_\max}$ to $L$, we take the average of (\ref{eq:qkLRenew}), by integrating each of the terms from $0$ to $1/2$ upon $\la$, using (\ref{eq:rpdip}) and (\ref{eq:terme2as}).
Adding the contributions coming from the two terms, we finally obtain for the dominant correction,
\beqa\label{eq:correctionRenew} 
L-\mean{X_\max}&\approx&
\frac{wc\,}{1-w}\left(\mathrm{B}_{\frac{1}{2}}(1-\th,1-\th)+\mathrm{B}_{\frac{1}{2}}(2-\th,-\th)\right)
L^{1-\th}
\nonumber\\
&=&\frac{wc\,}{1-w}\mathrm{B}_{\frac{1}{2}}\Big(1-\th,-\th\Big)L^{1-\th},
\eeqa
which has the same structure as (\ref{eq:correctionTdrpLt1}) or (\ref{eq:correctionzrplt1}).
The comment made below (\ref{eq:correctionTdrpGt1}) also holds here.
In the present case the expression in the right side of (\ref{eq:correctionRenew})
is proportional to 
$\mean{N_L}=w/(1-w)$ times the critical mean interval $\xmtd\sim L^{1-\th}$, as given in \cite{gl2001}.

\medskip\noindent 
$\bullet$ 
Likewise, for $\th>1$, the weight of the first term in (\ref{eq:qkLRenew}) dominates upon the second one, and we find for the correction of the mean to $L$,
\beq\label{eq:correctionRenew+}
L-\mean{X_\max}\approx \frac{w}{1-w}\xm,
\eeq
 showing that this correction is made of $\mean{N_L}=w/(1-w)$ intervals of size $\xm$.
This expression is therefore the perfect parallel of (\ref{eq:correctionTdrpGt1}) or (\ref{eq:correctionzrpgt1}).

\subsection{Probability for the last interval to be the longest one}

Lastly, we investigate the behaviour of ${Q_L}$ defined in (\ref{eq:defQ}) as the probability that $B_L$ is the longest interval.
An estimate of ${Q_L}$ for $w<1$ can be obtained by means of the inequality
\be
1-{Q_L}\lesssim\prob(B_L\le L/2).
\ee
In view of (\ref{eq:Blthalf}) and (\ref{eq:correctionRenew}), we infer that asymptotically for $L$ large, if $\th<1$, 
\be
{Q_L}\approx \frac{\mean{X_\max}}{L},
\ee
while, if $\th>1$, in view of (\ref{eq:Blthalf+}) and (\ref{eq:correctionRenew+}),
\be
1-Q_L\approx \frac{\th}{L}(L-\mean{X_\max}).
\ee
In other words, for $w<1$, $Q_L\to1$.
At criticality, $w=1$, $Q_L\to Q_\infty=0.626\dots$, if $\th<1$, while if $\th>1$, $Q_L\sim L^{-(1-1/\th)}$ \cite{gms2015}.
For $w>1$, $Q_L\to0$.

\begin{figure}[!ht]
\begin{center}
\includegraphics[angle=0,width=.8\linewidth]{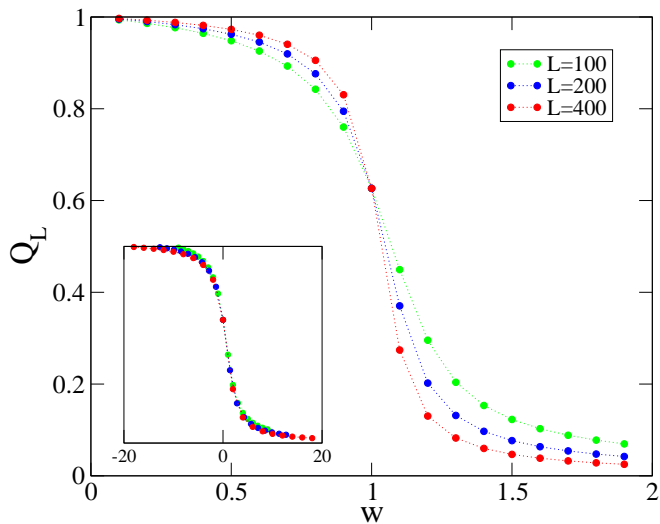}
\caption{Free renewal processes: probability ${Q_L}$ that the last interval, $B_L$, is the longest one,
for \textit{Example 1} (see (\ref{eq:ex1td})), for three different values of $L$.
The curves cross at the universal critical value $Q_{\infty}=0.626\dots$ for $w=1$.
Inset: after rescaling, $Q_L$ against the scaling variable $x=(w-1)L^{1/2}$.
}
\label{fig:Q}
\end{center}
\end{figure}

Figure \ref{fig:Q} depicts $Q_L$ as a function of $w$ for \textit{Example 1} (see (\ref{eq:ex1td})) and for three different sizes, crossing at the universal critical value $Q_{\infty}=0.626\dots$ for $w=1$ \cite{gms2015} and the data collapse obtained by using the scaling variable $x=(w-1)L^{1/2}$.

Table \ref {tab:renewal} summarises the results found in section \ref{sec:rpcondensed} and recapitulates the results for the two other phases (disordered and critical).
This table demonstrates a large degree of universality of the results, as was the case of table \ref{tab:tdrp}, with which it should be put in perspective.

\begin{table}[htbp]
\caption{Dominant asymptotic behaviours at large $L$ for free renewal processes with power-law distribution (\ref{eq:powerlaw}) for $f(k)$, in the different phases.
The results in columns 2 and 3 (critical phase) are taken from \cite {gl2001,gms2015}.
In the last column, $\xmtd\sim L^{1-\th}$ if $\th<1$, or $\xmtd\approx \mathrm{constant}$ if $\th>1$.}
\label{tab:renewal}
\begin{center}
\begin{tabular}{|c|c||c|c||c|}
\hline
&disordered&critical $\th<1$&critical $\th>1$&condensed\\
\hline
$\mean{N_L}$&$\frac{L}{\xm}$&$L^\th$&$\frac{L}{\xm}$&$\frac{w}{1-w}$\\
$\xmtd$&$\mathrm{constant}$&$ L^{1-\th}$&$\xm$&$L^{1-\th}$ or constant\\
$\mean{B_L}$&constant&$L$&$L^{2-\th}$&$L$\\
$\mean{X_\max}$&$ \ln L$&$ L $&$L^{1/\th}$&$ L$\\
$Z^{\free}(w,L)$&$\e^{L/\xi} $&$ 1 $&$1$&$ L^{-\th}$\\
\hline
\end{tabular}
\end{center}
\end{table}

\section{Conclusion}
\label{sec:conclusion}

Let us summarise the salient aspects of this study.

We first recalled the main features of the condensation transition taking place 
for random allocation models and \textsc{zrp}
in the thermodynamic limit ($L,n\to\infty$ with fixed ratio $\rho=L/n$), when the distribution of occupations is subexponential.
These occupations are independent and identically distributed random variables conditioned by the value of their sum.
The phase diagram is made of three phases: disordered, critical, and condensed.
The critical line $\rho=\rho_c(\th)$, where $\th>1$, separates the disordered phase at low density from the condensed phase at high density.
Condensation manifests itself by the occurrence, in the thermodynamic limit, of a unique site with macroscopic occupation.
In the language of particles and boxes (or sites), the condensate is by definition the site with the largest occupation.
In the language of sums of random variables used all throughout the present work, the condensate $X_{\max}$ is the unique summand with extensive value.
In the thermodynamic limit, the fraction $X_\max/L$ no longer fluctuates and takes the asymptotic value $1-\rho_c/\rho$.

A second scenario for the same class of models consists in taking the $L\to\infty$ limit keeping the number of sites (or summands) fixed.
In this limit there is again a single extensive summand $X_\max$, but now the fraction $X_\max/L$ tends to unity, which means that condensation is \textit{total}.
The novelty is that this occurs irrespective of the existence of a first moment $\xm$, or in other words, irrespective of whether $\th$ is smaller or larger than one.
If $L$ is large but finite, the distribution of $X_\max$ is peaked, with a width $L-\mean{X_\max}$ scaling as 
$L^{1-\th}$ if $\th<1$, with a known amplitude, or asymptotically equal to $(n-1)\xm$, if $\th>1$.
Note that, in contrast to the previous case, one can no longer speak of a phase transition, nor even of a phase, since the system is made of a finite number $n$ of summands (or sites).

This scenario is a good preparation for the study of condensation in free and tied-down renewal processes, with power-law distribution of intervals (\ref{eq:powerlaw}), which is the main motivation of the present work.
Instead of particle occupations and sites one speaks in terms of renewal events and intervals, whose sizes sum up to a fixed value $L$.
The novelty---and complication---is that the number of these renewal events, or equivalently of intervals, $N_L$, fluctuates.
For instance these renewal points are the passages by the origin of a random walk, as depicted in figure \ref{fig:tdrw}.
A weight $w$ is attached to each renewal event.
In the language of random walks (or of polymer chains) $w$ represents the reward or penalty when the walk touches the origin \cite{fisher,gia1,gia2}.
A high value of $w$ favours configurations with a large number of intervals $N_L$, i.e., a disordered phase---or localised phase in the language of random walks.
A low value of $w$ favours configurations with a small number of intervals $N_L$, i.e., a condensed phase---or delocalised phase in the language of random walks.
It is therefore intuitively clear that the same scenario of total condensation as seen above should prevail,
where now the driving force is no longer a change in the density, $\rho$, but a change in the value of the weight $w$ attached to each interval (or summand).
In this respect it is worth noting the similarity between equations (\ref{eq:correctionzrplt1}), (\ref{eq:correctionTdrpLt1}) and (\ref{eq:correctionRenew}) on one hand, and the similarity between equations (\ref{eq:correctionzrpgt1}), (\ref{eq:correctionTdrpGt1}) and (\ref{eq:correctionRenew+}) on the other hand.

It turns out that, in the condensed phase, when $L\to\infty$, the distribution of the number of intervals, $N_L$, is superuniversal, i.e., model independent, since it only depends on $w$ and not even on the index $\th$ of the power-law decay (\ref{eq:powerlaw}).
This distribution is geometric for free renewal processes, while it is the convolution of two such distributions for \textsc{tdrp}.
More generally, an important distinction is to be made according to whether $\th$ is less or larger than unity.
In the first case the distribution $f(k)$ has no first moment, atypical events play a major role and the system becomes self-similar at criticality.
In the second case the observables of interest depend on the first moment $\xm$, which is finite.

In closing, let us broaden the perspective.
The phase transition occurring when $w$ passes through unity is second order for the density of intervals $\nu$ (defined in (\ref{eq:defnu})) if $\th<1$, and first order if $\th>1$.
On the other hand the correlation length diverges at the transition, see (\ref{eq:xi}).
The transition is therefore mixed order as was pointed out for the particular case of \textsc{tdrp} with \textit{Example 2} (see (\ref{eq:fkex2})) in \cite{burda3,bar2}.
Furthermore, the magnetisation, defined as the alternating sum $m=(X_1-X_2+X_3-\cdots)/L$,
changes, when $L\to\infty$, from the value $0$ in the disordered phase to $\pm1$ in the condensed phase since condensation is total.
More on this can be found in \cite{bar2}.
If $\th<1$ the distribution of the magnetisation at criticality is broad and self-similar,
both for free \cite{lamperti58,gl2001} and tied-down renewal processes \cite{wendel2}.
At criticality, for $\th<1$, the non stationary two-time (or two-space) correlation function is also self-similar, again for both processes \cite{gl2001,wendel2}.%
\footnote{After submission of the present work, a study devoted to the statistics of $X_\max$ in the range $(L/2,L)$ for tied-down or free renewal processes at criticality ($w=1$) was presented in \cite{barkai}.
For the \textsc{tdrp} case the result (\ref{eq:tdc}) with $w=1$ is obtained.
For the free renewal case, \cite{barkai} predicts, if $w=1$,
\be
q(k|L)=g(k)(\mean{N^{\free}_{L-k}}-\mean{N^{\free}_{L-k-1}})+f(k)(1+\mean{N^{\free}_{L-k}}),
\ee
which is (\ref{eq:beautiful}), with $w=1$, noting that $Z^{\td}(1,L)=\mean{N^{\free}_L}-\mean{N^{\free}_{L-1}}$, as is clear by taking the generating functions of both sides.}

\begin{acknowledgements}
It is a pleasure to thank G Giacomin, M Loulakis and J-M Luck for enlightening discussions.
I am also indebted to S Grosskinsky and S Janson for useful correspondence.
\end{acknowledgements}

\section*{Appendix}

\appendix

\section{On equation (\ref{eq:FnLsur2})}
\label{app:heuristic}

Let us explain the argument leading to (\ref{eq:FnLsur2}) and the origin of the hierarchical structure mentioned in \S\ref{sec:details}%
\footnote{I am indebted to M Loulakis for sharing his comments on this part with me.}.

\begin{enumerate}
\item In the uphill region where $L-X_\max$ is finite, $X_\max$ is the unique big summand and (\ref{eq:uphill-subexp}) holds.
This property stems from the fact that when the sum of $n$ subexponential random variables is conditioned to a large value $L$, all the dependence is absorbed by the maximum and the ensemble of $n-1$ smaller variables becomes asymptotically independent.
This property, initially put forward by early workers, has been progressively refined in subsequent studies \cite{ferrari,armendariz2011,janson}.

\item In the dip region, where $X_\max>L/2$, since $L-X_\max$ gets large, the sum 
$\sum_{i=1}^{n-1}X_i$ becomes subjected to a large deviation event.
This event will be realised by $X^{(2)}$, the second largest summand, typically equal to $L-X_\max$.
We thus obtain (\ref{eq:zrpMaxDip}).
\item One can now iterate the reasoning.
If $X_\max=k\le L/2$, the difference $L-k\ge k$ cannot accommodate a single big summand $X^{(2)}=j$ since the latter should be less than $X_\max$.
Now
\be
L-X_\max-X^{(2)}\stackunder{L\to\infty}{\approx}\sum_{i=1}^{n-2}X_i.
\ee
where the sum in the right side is subjected to a large deviation, which will be realised by a third large summand $X^{(3)}$.
Since $L-k-j$ should be less than $j$, the constraint $j\ge (L-k)/2$ holds.
Moreover $X^{(2)}\le X_\max$ imposes the condition $(L-k)/2\le k$, i.e. $k\ge L/3$.
We are thus lead to the asymptotic estimate
\be
\prob(X_\max=k|S_n=L)\approx \sum_{j=\frac{L-k}{2}}^{k}n(n-1)\frac{f(k)f(j)Z_{n-2}(L-k-j)}{Z_n(L)},
\ee
and therefore
\beq\label{app:ansatz}
\prob(L/3\le X_\max\le L/2|S_n=L)\comport{\approx}{}{} \sum_{k=L/3}^{L/2}\
\sum_{j=\frac{L-k}{2}}^k
n(n-1)\frac{f(k)f(j)Z_{n-2}(L-k-j)}{Z_n(L)}.
\eeq

\end{enumerate}

For $\th<1$, the analysis of this expression in the continuum limit leads to 
\beq\label{app:Pn-teta}
\prob(L/3\le X_\max\le L/2|S_n=L)\approx (n-1)(n-2)c^2\frac{A(\th)}{L^{2\th}},
\eeq
where the amplitude $A(\th)$ is given by
\beq\label{app:amplit}
A(\th)=\int_{1/3}^{1/2}\dd x\,
\int_{\frac{1-x}{2}}^{x}\dd y\,\frac{1}{[xy(1-x-y)]^{1+\th}}.
\eeq
For instance, $A(1/2)=2\pi$, $A(1/3)=9\sqrt{3}\,\Gamma(2/3)^3/(4\pi)$.

For $\th>1$, the analysis of (\ref{app:ansatz}) yields
\beq\label{eq:gt1L3L2C}
\prob(L/3<X_\max<L/2|S_n=L)\approx \frac{(n-1)c\,2^{2+2\th}}{L^{1+\th}} \frac{(n-2)\xm}{2},
\eeq
for continuous random variables, and
\beq\label{eq:gt1L3L2D}
\prob(L/3\le X_\max\le L/2|S_n=L)\approx \frac{(n-1)c\,2^{2+2\th}}{L^{1+\th}} \frac{(n-2)\xm+1}{2},
\eeq
for discrete ones.

One can iterate the reasoning leading to (\ref{app:ansatz}) and derive the weights of the successive sectors $(L/4,L/3)$, $(L/5,L/4$), etc.
For instance, for $X_\max$ in the interval $(L/4,L/3)$, one finds
\be
\prob(X_\max=k|S_n=L)\approx \sum_{j=\frac{L-k}{3}}^{k}\sum_{i=\frac{L-k-j}{2}}^jn(n-1)(n-2)\frac{f(k)f(j)f(i)Z_{n-3}(L-k-j-i)}{Z_n(L)},
\ee
which yields
\beq
\prob(L/4\le X_\max\le L/3|S_n=L)\sim L^{-\gamma},\qquad 
\gamma= \left\{
\begin{array}{ll}
3\th \ & \textrm{if } \th\le2
\vspace{4pt}\\
2(1+\th)\ & \textrm{if }\th>2,
\end{array}
\right.
\eeq 
For example, if $\th=1/2$, one finds
\be
\prob(L/4\le X_\max\le L/3|S_n=L)\approx (n-1)(n-2)(n-3)c^3\frac{\sqrt{2}\,\pi/3}{L^{3/2}}.
\ee

\section{Weight of the maximum in the left region for a L\'evy $\frac{1}{2}$ stable law}
\label{app:levy}

We want to determine the weight of the maximum in the left region considered in \S \ref{sec:details},
\be
P_n=\prob(X_\max\le L/2|S_n=L)=\sum_{k=0}^{L/2}p_n(k|L)=1-\sum_{k=L/2+1}^{L}p_n(k|L)=1-\sum_{k=L/2+1}^{L}n\pi_n(k|L),
\ee
on the particular example of a a L\'evy $\frac{1}{2}$ stable law.
We use a continuum formalism where distributions are densities and variables are real numbers
for the particular case where $f(k)$ is the L\'evy $\frac{1}{2}$ stable density (\ref{eq:stable12Lev}),
\be
f(k)=\mathcal{L}_{\frac{1}{2},c}(k)=\frac{c}{k^{3/2}}\e^{-\pi c^2/k}.
\ee
Likewise, considering $L$ as a real number and $Z_n(L)$ as a density,
\be
Z_n(L)=\mathcal{L}_{\frac{1}{2},nc}(k)=\frac{nc}{L^{3/2}}\e^{-n^2\pi c^2/L}.
\ee
Thus
\be
\pi_n(k|L)=\frac{f(k)Z_{n-1}(L-k)}{Z_n(L)}
\ee
is explicit.
Setting $k=L/t$, we obtain
\be
P_n=1-\frac{(n-1)c}{\sqrt{L}}\int_{1}^2\dd t\,\frac{t\,\e^{-\frac{\pi c^2(n-t)^2}{L(t-1)}}}{(t-1)^{3/2}}.
\ee
Setting $c/\sqrt{L}=b/(2\sqrt{\pi})$, we finally get
\be
P_n=\frac{1}{2}\Bigg(2-n\erfc\Big(\frac{(n-2)b}{2}\Big)+(n-2)\e^{(n-1)b^2}\erfc\Big(\frac{nb}{2}\Big)
\Bigg).
\ee
For $L$ large, expanding in powers of $b$, we obtain
\bea
P_n
&=&(n-1)(n-2)\Big(\frac{b^2}{2}-n\frac{b^3}{3\sqrt{\pi}}+(n-1)\frac{b^4}{4}+\cdots\Big)
\nonumber\\
&=&(n-1)(n-2)\Big(2\pi\frac{c^2}{L}-\frac{8\pi n}{3}\frac{c^3}{L^{3/2}}+(n-1)4\pi^2\frac{c^4}{L^2}+\cdots\Big).
\eea
For \textit{Example 1}, $c=1/(2\sqrt{\pi})$ (see (\ref{eq:ex1td})), 
the first term of the expansion,
\be
P_n\approx \frac{(n-1)(n-2)}{2L},
\ee
matches the predictions made in (\ref{app:Pn-teta}) and (\ref{app:amplit}).

\end{document}